\documentclass[12pt]{article}
\usepackage{graphicx} %macropackage to manage figures
\usepackage[spanish, english]{babel} % to have labels in spanish
\usepackage{float} %interface for floating object
\usepackage[utf8]{inputenc} %allow utf-8 input
\usepackage[T1]{fontenc} %use 8-bit T1 fonts
\usepackage{outline}
\usepackage{pmgraph}
\usepackage[normalem]{ulem}
\usepackage{amsmath,mathdots}
\usepackage{amsfonts,amssymb,amsthm,yhmath}
\usepackage{fancyhdr}
\usepackage{vmargin}
\usepackage{tikz}
\usepackage{physics}
\usepackage{cancel}
\usepackage{bm}
\usepackage{hyperref}
\usepackage[scaled]{helvet}
\usepackage{color}
\usepackage{siunitx}
\usepackage{array}
\usepackage{gensymb}
\usepackage{tabularx}
\usepackage{extarrows}
\usepackage{booktabs}
\usepackage[export]{adjustbox}
\usepackage{braket}
\usetikzlibrary{fadings}
\usetikzlibrary{patterns}
\usetikzlibrary{shadows.blur}
\usetikzlibrary{shapes}
\usepackage{caption}
\usepackage{subcaption}

\decimalpoint
\hypersetup{
    colorlinks=true,
    linkcolor=black,
    filecolor=magenta,
    urlcolor=blue,
}

\title{Cálculo de propiedades ópticas de metamateriales\footnote{Curso
    presentado en la XXVIII Escuela de Verano en Física, ICF- e
    IF-UNAM, Cuernavaca, Mor. y Cd de México,
  junio 21-julio 2, 2021}}
\author{ Merlyn Jaqueline Juárez-Gutiérrez y W. Luis Mochán\\
  Instituto de Ciencias Físicas}

\begin{document}
\selectlanguage{spanish}
\maketitle
\selectlanguage{english}
\begin{abstract}
  We present an introduction to metamaterials, some of their optical
  properties, and examples of their uses. We develop an efficient
  theory for the calculation of the macroscopic permittivity of binary
  systems and systems with more components, in the non-retarded case and in the
  general case, and we present its implementation in a computational
  package and illustrate its use. We discuss some applications regarding
  the design of optimized nanostructured optical devices and we
  discuss the linear and non-linear properties obtained.
\end{abstract}

\selectlanguage{spanish}
\begin{abstract}
  Presentamos una introducción a los metamateriales, algunas de sus
  propiedades ópticas y ejemplos de sus usos. Desarrollamos una teoría
  eficiente para el cálculo de su permitividad macroscópica en
  sistemas binarios o con más componentes, en el caso no retardado y
  en el caso general, y presentamos su implementación en un paquete
  computacional y su uso. Finalizamos discutiendo algunas aplicaciones
  del mismo para el diseño de dispositivos ópticos nanoestructurados
  optimizados y discutimos las propiedades lineales y no lineales
  obtenidas.
\end{abstract}
\section{Introducción}

Un metamaterial es un material artificial formado por
dos o más materiales alternados. Los metamateriales se pueden definir como
{\em un arreglo de elementos estructurales artificiales, diseñados
  para alcanzar propiedades electromagnéticas ventajosas e
  inusuales}\cite{Metamorphose}, de acuerdo al
{\em Virtual Institute for Artificial Electromagnetic Materials and
  Meta-Materials}.
Dichas propiedades están
determinadas por sus constituyentes básicos, a los que se denomina
ocasionalmente como {\em meta-átomos}, los cuales son objetos hechos de
materiales usuales, así como por su forma y disposición, y pueden ser
muy distintas a las de los materiales que los conforman, llegando a
a ser muy exóticas. Pueden ser diseñadas y entonadas escogiendo las formas,
estructuras internas, tamaños, orientaciones mutuas, etc., de sus
meta-átomos.  Sus funciones repuesta pueden ser
modificadas mediante señales externas e internas y pueden ser
controladas mediante microprocesadores
programables.\cite{IntroductiontoMetamaterialsandNanophotonics}

\subsection{Materiales plasmónicos}
\begin{figure}
\centering
\input{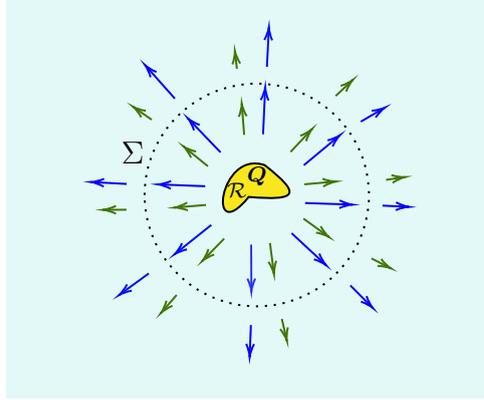}
\caption{Acumulación de carga $Q$ en cierta región $\mathcal{R}$
  dentro de un conductor homogéneo, rodeada por una superficie
  $\Sigma$ sobre la cual se aplica la ley de Gauss al campo producido
  por $Q$, campo que produce una densidad de corriente que fluye a
  través de $\Sigma$. La corriente modifica $Q$ conforme transcurre el
  tiempo $t$ produciendo oscilaciones de plasma.}
\label{Bulkplasmon}
\end{figure}

\begin{figure}
  \centering
  \begin{tikzpicture}[x=0.9pt,y=0.9pt,yscale=-1,xscale=1]
%uncomment if require: \path (0,193); %set diagram left start at 0,
%and has height of 193

%Shape: Inductor (Air Core) [id:dp34048340969550106]
\draw (119.21,42.46) -- (119.25,59.21) .. controls (131.85,59.42) and
(142.62,61.72) .. (146.39,65.01) .. controls (150.17,68.29) and
(146.17,71.88) .. (136.32,74.06) .. controls (128.64,75.74) and
(118.71,76.44) .. (109.06,75.98) .. controls (105.3,75.99) and
(102.24,75.16) .. (102.24,74.14) .. controls (102.24,73.11) and
(105.29,72.27) .. (109.05,72.26) .. controls (118.7,71.76) and
(128.63,72.42) .. (136.32,74.06) .. controls (144.52,75.98) and
(149.16,78.66) .. (149.17,81.48) .. controls (149.18,84.29) and
(144.54,87) .. (136.36,88.95) .. controls (128.68,90.63) and
(118.74,91.33) .. (109.09,90.87) .. controls (105.33,90.88) and
(102.27,90.05) .. (102.27,89.03) .. controls (102.27,88) and
(105.32,87.16) .. (109.08,87.15) .. controls (118.73,86.65) and
(128.67,87.3) .. (136.36,88.95) .. controls (144.55,90.86) and
(149.2,93.55) .. (149.2,96.36) .. controls (149.21,99.18) and
(144.57,101.88) .. (136.39,103.84) .. controls (128.71,105.52) and
(118.77,106.22) .. (109.13,105.76) .. controls (105.36,105.77) and
(102.31,104.94) .. (102.31,103.91) .. controls (102.3,102.89) and
(105.36,102.05) .. (109.12,102.04) .. controls (118.76,101.54) and
(128.7,102.19) .. (136.39,103.84) .. controls (146.25,105.97) and
(150.26,109.55) .. (146.5,112.85) .. controls (142.74,116.15) and
(131.98,118.49) .. (119.38,118.76) -- (119.42,135.51) ;
%Shape: Rectangle [id:dp47577099884182095]
\draw [fill={rgb, 255:red, 245; green, 166; blue, 35 } ,fill opacity=1
] (87,27.5) -- (151.17,27.5) -- (151.17,43) -- (87,43) -- cycle ;
%Shape: Circle [id:dp24543618839132486]
\draw [fill={rgb, 255:red, 126; green, 211; blue, 33 } ,fill opacity=1
] (109.67,145.26) .. controls (109.67,139.88) and (114.04,135.51)
.. (119.42,135.51) .. controls (124.8,135.51) and (129.17,139.88)
.. (129.17,145.26) .. controls (129.17,150.65) and (124.8,155.01)
.. (119.42,155.01) .. controls (114.04,155.01) and (109.67,150.65)
.. (109.67,145.26) -- cycle ;
%Shape: Inductor (Air Core) [id:dp8275323218505621]
\draw (42.21,42.46) -- (42.23,51.56) .. controls (54.83,51.66) and
(65.6,52.9) .. (69.37,54.68) .. controls (73.14,56.46) and
(69.14,58.41) .. (59.29,59.61) .. controls (51.61,60.53) and
(41.67,60.92) .. (32.02,60.68) .. controls (28.26,60.69) and
(25.21,60.24) .. (25.21,59.68) .. controls (25.2,59.13) and
(28.26,58.67) .. (32.02,58.66) .. controls (41.67,58.37) and
(51.6,58.72) .. (59.29,59.61) .. controls (67.48,60.64) and
(72.12,62.09) .. (72.13,63.62) .. controls (72.13,65.15) and
(67.49,66.62) .. (59.31,67.69) .. controls (51.63,68.61) and
(41.69,69) .. (32.04,68.76) .. controls (28.28,68.77) and
(25.23,68.33) .. (25.22,67.77) .. controls (25.22,67.21) and
(28.28,66.75) .. (32.04,66.74) .. controls (41.68,66.46) and
(51.62,66.81) .. (59.31,67.69) .. controls (67.5,68.72) and
(72.14,70.18) .. (72.15,71.71) .. controls (72.15,73.24) and
(67.51,74.71) .. (59.33,75.78) .. controls (51.64,76.7) and
(41.71,77.09) .. (32.06,76.85) .. controls (28.3,76.86) and
(25.24,76.41) .. (25.24,75.86) .. controls (25.24,75.3) and
(28.29,74.84) .. (32.06,74.83) .. controls (41.7,74.55) and
(51.64,74.89) .. (59.33,75.78) .. controls (69.18,76.93) and
(73.19,78.86) .. (69.43,80.66) .. controls (65.67,82.46) and
(54.9,83.74) .. (42.3,83.9) -- (42.32,93) ;
%Shape: Rectangle [id:dp7392561467625306]
\draw [fill={rgb, 255:red, 245; green, 166; blue, 35 } ,fill opacity=1
] (10,27.5) -- (74.17,27.5) -- (74.17,43) -- (10,43) -- cycle ;
%Shape: Circle [id:dp15959587945734588]
\draw [fill={rgb, 255:red, 126; green, 211; blue, 33 } ,fill opacity=1
] (32.57,102.75) .. controls (32.57,97.36) and (36.94,93)
.. (42.32,93) .. controls (47.71,93) and (52.07,97.36)
.. (52.07,102.75) .. controls (52.07,108.13) and (47.71,112.5)
.. (42.32,112.5) .. controls (36.94,112.5) and (32.57,108.13)
.. (32.57,102.75) -- cycle ;

%Straight Lines [id:da8898992399758622]
\draw (95.32,102.75) -- (95.17,147) ; \draw [shift={(95.17,147)},
  rotate = 270.2] [color={rgb, 255:red, 0; green, 0; blue, 0 } ][line
  width=0.75] (0,5.59) -- (0,-5.59) ; \draw [shift={(95.32,102.75)},
  rotate = 270.2] [color={rgb, 255:red, 0; green, 0; blue, 0 } ][line
  width=0.75] (0,5.59) -- (0,-5.59) ;

% Text Node
\draw (5,58.4) node [anchor=north west][inner sep=0.75pt]    {$k$};
% Text Node
\draw (57,100.4) node [anchor=north west][inner sep=0.75pt]    {$m$};
% Text Node
\draw (132,134.4) node [anchor=north west][inner sep=0.75pt]    {$m$};
% Text Node
\draw (83,113.4) node [anchor=north west][inner sep=0.75pt]    {$y$};

\end{tikzpicture}
  \caption{\label{OscArmonicoSimple} Oscilador armónico de masa $m$ y
    constante $k$ en su posición de equilibrio y desplazado una
    distancia $y$.}
\end{figure}
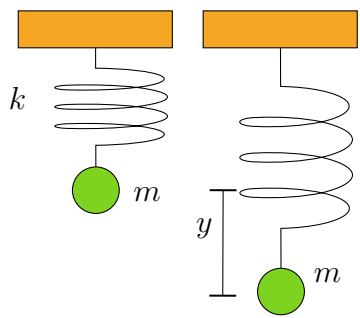

Si un metamaterial tiene componentes metálicos, estos
pueden presentar resonancias asociadas a los movimientos oscilatorios
colectivos de sus electrones de conducción, denominados de acuerdo a sus
características como plasmones de bulto, de superficie o
localizados. La frecuencia de estos movimientos en un sistema infinito se
denomina como
frecuencia de plasma $\omega_{p}$. Para estimarla,
considere el modelo más simple de un conductor, un gas de electrones
que en equilibrio tienen una densidad de número $n_{0}$, y que son libres de
moverse en un entorno positivamente cargado, de manera que el sistema
en equilibrio sea neutro. Si debido a alguna compresión o rarefacción
del gas de electrones se produjera una acumulación de carga $Q$
localizada en alguna región $\mathcal{R}$, ésta produciría un campo
eléctrico $\bm{E} (\bm{r})$, como ilustra la figura
\ref{Bulkplasmon}. De acuerdo a la segunda ley de Newton, los
electrones adquirirían una aceleración
$d^{2}\bm{r}/dt^{2} = -e\bm{E}(\bm{r},t)/m$, donde $m$ y $-e$
son la masa y la carga eléctrica. La velocidad adquirida
por los electrones resultaría en una corriente eléctrica
$\bm{j}(\bm{r},t) = -n_0 e \bm v=-n_{0}e({d\bm{r}}/{dt})$ que obedece
la ecuación de movimiento,
${\partial \bm{j}(\bm{r},t)}/{\partial t}=
-\frac{n_{0}e^{2}}{m}\bm{E}(\bm{r},t)$. Integrando la ecuación
diferencial de la corriente sobre una superficie $\Sigma$ que rodee
completamente la carga y usando la ecuación de continuidad y la Ley de
Gauss para el campo eléctrico obtenemos una ecuación diferencial para
la carga $Q$ encerrada en $\Sigma$,
\begin{equation}
  \label{ChargeDifEq}
  \frac{d^{2}Q}{dt^{2}}=-\frac{4\pi n_0 e^{2}}{m}Q,
\end{equation}
la cual es una ecuación diferencial idéntica a la de un oscilador
armónico simple como el que se ilustra en la fig.
\ref{OscArmonicoSimple}.
Sustituyendo la Ley de Hooke $F= -ky$ en la segunda ley de Newton
$ma=F$ para un oscilador con constante $k$ y masa $m$, obtenemos \cite{plasmons}
\begin{equation}
  \label{OAEq}
  \frac{d^{2}y}{dt^{2}}=-\omega^2y,
\end{equation}
donde $\omega=\sqrt{k/m}$, y cuya solución,
$y(t)=y_{0}\cos(\omega t+\phi_0)$ es un movimiento periódico con una
frecuencia $\omega$ que depende de $k$ y $m$.  Comparando las ecs.
\eqref{ChargeDifEq} y \eqref{OAEq} notamos que puede existir carga en
el seno de nuestro metal modelo, pero ésta oscilaría con la {\em
  frecuencia de plasma} $\omega_p$ dada por
\begin{equation}
  \label{plasmafec}
  \omega _{p}^{2} = \frac{4\pi n_0e^{2}}{m}.
\end{equation}
La repulsión mutua entre electrones los impulsa lejos de regiones en
que haya una densidad electrónica excedente, por arriba de su valor
nominal. El movimiento consecuente prosigue aun después de que el
sistema se neutraliza debido a la inercia electrónica, que los hace
proseguir su camino hasta que en la región original disminuye tanto la
densidad de electrones que aparezca una carga neta positiva que frena
a los electrones en fuga y los hace regresar, hasta que su repulsión
mutua los frena, habiendo regresado a la configuración inicial. Este
proceso se repite periódicamente y su frecuencia $\omega_p$ está
relacionada con la repulsión coulombiana, proporcional a $e^2$, la
densidad de número electrónica $n_0$ y la inercia electrónica
caracterizada por $m$.

\begin{figure}
  \centering
  \input{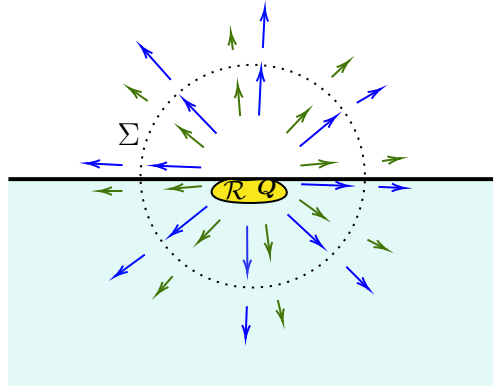}
  \caption{Región $\mathcal{R}$ en la superficie de un conductor
    semi-infinito en la que hay una carga $Q$, la cual produce un
    campo eléctrico $\bm{E}(\bm{r},t)$ (líneas azules) y éste a su vez
    produce una densidad de corriente (líneas verdes) que atraviesan
    aquella parte de la superficie lejana $\Sigma$ que se halla dentro
    del conductor.}
\label{Surfplasmon}
\end{figure}

En lugar de un medio infinito, consideremos ahora un medio
semiinfinito separado del vacío por una superficie plana.  Un análisis
análogo nos permite obtener la frecuencia del {\em plasmón de
  superficie}, considerando ahora un exceso de carga $Q$ en una región
$\mathcal{R}$ en la interfaz, como ilustra la fig.  \ref{Surfplasmon}, el cual
produce un campo eléctrico $\bm{E}(\bm{r},t)$ que induce corrientes en
el conductor.

Usando la ecuación dinámica de la densidad de corriente podemos
escribir la ecuación dinámica de la carga como hicimos en el caso del
plasmón de bulto, con la diferencia que la densidad de corriente en
este caso sólo fluye a través de la mitad $\Sigma/2$ de la superficie $\Sigma $
que se halla en el interior del metal,
\begin{equation}
  \begin{split}
    \frac{d^{2}}{dt^{2}}Q = & -\int_{\Sigma}d\bm{a}\cdot\frac{\partial}{\partial t}\bm{j},\\
    = & ne \int_{\Sigma/2}d\bm{a}\cdot\frac{\partial}{\partial t}\bm{v},\\
    = & \frac{1}{2}\frac{ne^{2}}{m}\int_{\Sigma}d\bm{a}\cdot \bm{E}, \\
    = & -\frac{2\pi ne^{2}}{m}Q.
  \end{split}
\end{equation}
Comparando esta ecuación con la ec. \eqref{OAEq} identificamos la
frecuencia del {\em plasmón de superficie} $\omega_{\text{sp}}$, dada
por
\begin{equation}
  \omega_{sp}^{2}=\frac{2\pi ne^{2}}{m} = \frac{\omega_{p}^{2}}{2}.
  \label{surfaceplasmonfrecuency}
\end{equation}

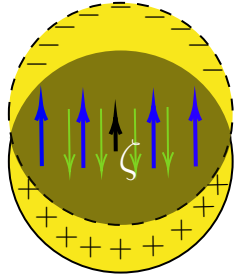
\begin{figure}
  \centering
  \tikzset{every picture/.style={line width=0.75pt}} %set default line width to 0.75pt

\begin{tikzpicture}[x=0.75pt,y=0.75pt,yscale=-1,xscale=1]
%Shape: Circle [id:dp14966767726976038]
\draw [fill={rgb, 255:red, 248; green, 231; blue, 28 } ,fill opacity=1
] (219.75,132.08) .. controls (219.75,101.2) and (244.78,76.17)
.. (275.67,76.17) .. controls (306.55,76.17) and (331.58,101.2)
.. (331.58,132.08) .. controls (331.58,162.97) and (306.55,188)
.. (275.67,188) .. controls (244.78,188) and (219.75,162.97)
.. (219.75,132.08) -- cycle ;
%Shape: Circle [id:dp9677285416132982]
\draw [fill={rgb, 255:red, 248; green, 231; blue, 28 } ,fill opacity=1
][dash pattern={on 4.5pt off 4.5pt}] (219.75,108.08) .. controls
(219.75,77.2) and (244.78,52.17) .. (275.67,52.17) .. controls
(306.55,52.17) and (331.58,77.2) .. (331.58,108.08) .. controls
(331.58,138.97) and (306.55,164) .. (275.67,164) .. controls
(244.78,164) and (219.75,138.97) .. (219.75,108.08) -- cycle ;
%Shape: Path Data [id:dp41415328569096044]
\draw [draw opacity=0][fill={rgb, 255:red, 25; green, 23; blue, 2 }
  ,fill opacity=0.53 ] (275.67,164) .. controls (248.27,164) and
(225.42,145.14) .. (220.17,120.08) .. controls (225.42,95.03) and
(248.27,76.17) .. (275.67,76.17) .. controls (303.06,76.17) and
(325.91,95.03) .. (331.17,120.08) .. controls (325.91,145.14) and
(303.06,164) .. (275.67,164) -- cycle ;

%Straight Lines [id:da9936815684711549]
\draw [line width=1.5] (273.03,112.79) -- (273.17,126.96) ; \draw
      [shift={(273,109.79)}, rotate = 89.44] [color={rgb, 255:red, 0;
          green, 0; blue, 0 } ][line width=1.5] (8.53,-2.57)
      .. controls (5.42,-1.09) and (2.58,-0.23) .. (0,0) .. controls
      (2.58,0.23) and (5.42,1.09) .. (8.53,2.57) ;
%Straight Lines [id:da577848764958209]
\draw [color={rgb, 255:red, 0; green, 4; blue, 243 } ,draw opacity=1
][line width=1.5] (235.87,105.83) -- (236.25,134.39) ; \draw
      [shift={(235.83,102.83)}, rotate = 89.24] [color={rgb, 255:red,
          0; green, 4; blue, 243 } ,draw opacity=1 ][line width=1.5]
      (8.53,-2.57) .. controls (5.42,-1.09) and (2.58,-0.23) .. (0,0)
      .. controls (2.58,0.23) and (5.42,1.09) .. (8.53,2.57) ;
%Straight Lines [id:da23709733272432076]
\draw [color={rgb, 255:red, 40; green, 0; blue, 255 } ,draw opacity=1
][line width=1.5] (312.79,105.99) -- (313.17,134.55) ; \draw
      [shift={(312.75,102.99)}, rotate = 89.24] [color={rgb, 255:red,
          40; green, 0; blue, 255 } ,draw opacity=1 ][line width=1.5]
      (8.53,-2.57) .. controls (5.42,-1.09) and (2.58,-0.23) .. (0,0)
      .. controls (2.58,0.23) and (5.42,1.09) .. (8.53,2.57) ;
%Straight Lines [id:da356188359843302]
\draw [color={rgb, 255:red, 22; green, 0; blue, 255 } ,draw opacity=1
][line width=1.5] (256.85,106.76) -- (257,135.32) ; \draw
      [shift={(256.83,103.76)}, rotate = 89.7] [color={rgb, 255:red,
          22; green, 0; blue, 255 } ,draw opacity=1 ][line width=1.5]
      (8.53,-2.57) .. controls (5.42,-1.09) and (2.58,-0.23) .. (0,0)
      .. controls (2.58,0.23) and (5.42,1.09) .. (8.53,2.57) ;
%Straight Lines [id:da517585674881536]
\draw [color={rgb, 255:red, 22; green, 0; blue, 255 } ,draw opacity=1
][line width=1.5] (292.02,106.91) -- (292.17,135.47) ; \draw
      [shift={(292,103.91)}, rotate = 89.7] [color={rgb, 255:red, 22;
          green, 0; blue, 255 } ,draw opacity=1 ][line width=1.5]
      (8.53,-2.57) .. controls (5.42,-1.09) and (2.58,-0.23) .. (0,0)
      .. controls (2.58,0.23) and (5.42,1.09) .. (8.53,2.57) ;
%Straight Lines [id:da6814351373246353]
\draw [color={rgb, 255:red, 126; green, 211; blue, 33 } ,draw
  opacity=1 ][line width=0.75] (283.14,134.98) -- (282.7,105.43) ;
\draw [shift={(283.17,136.98)}, rotate = 269.15] [color={rgb, 255:red,
    126; green, 211; blue, 33 } ,draw opacity=1 ][line width=0.75]
(8.74,-2.63) .. controls (5.56,-1.12) and (2.65,-0.24) .. (0,0)
.. controls (2.65,0.24) and (5.56,1.12) .. (8.74,2.63) ;
%Straight Lines [id:da19196139788350697]
\draw [color={rgb, 255:red, 126; green, 211; blue, 33 } ,draw
  opacity=1 ][line width=0.75] (250.16,135.21) -- (249.72,105.65) ;
\draw [shift={(250.19,137.21)}, rotate = 269.15] [color={rgb, 255:red,
    126; green, 211; blue, 33 } ,draw opacity=1 ][line width=0.75]
(8.74,-2.63) .. controls (5.56,-1.12) and (2.65,-0.24) .. (0,0)
.. controls (2.65,0.24) and (5.56,1.12) .. (8.74,2.63) ;
%Straight Lines [id:da7557269004865166]
\draw [color={rgb, 255:red, 126; green, 211; blue, 33 } ,draw
  opacity=1 ][line width=0.75] (299.33,134.18) -- (299.06,104.62) ;
\draw [shift={(299.35,136.18)}, rotate = 269.48] [color={rgb, 255:red,
    126; green, 211; blue, 33 } ,draw opacity=1 ][line width=0.75]
(8.74,-2.63) .. controls (5.56,-1.12) and (2.65,-0.24) .. (0,0)
.. controls (2.65,0.24) and (5.56,1.12) .. (8.74,2.63) ;
%Straight Lines [id:da6683208735835106]
\draw [color={rgb, 255:red, 126; green, 211; blue, 33 } ,draw
  opacity=1 ][line width=0.75] (266.08,135.02) -- (265.81,105.47) ;
\draw [shift={(266.1,137.02)}, rotate = 269.48] [color={rgb, 255:red,
    126; green, 211; blue, 33 } ,draw opacity=1 ][line width=0.75]
(8.74,-2.63) .. controls (5.56,-1.12) and (2.65,-0.24) .. (0,0)
.. controls (2.65,0.24) and (5.56,1.12) .. (8.74,2.63) ;

% Text Node
\draw (268,47.4) node [anchor=north west][inner sep=0.75pt] {$-$};
% Text Node
\draw (285,52.4) node [anchor=north west][inner sep=0.75pt] {$-$};
% Text Node
\draw (298,58.4) node [anchor=north west][inner sep=0.75pt] {$-$};
% Text Node
\draw (307,66.4) node [anchor=north west][inner sep=0.75pt] {$-$};
% Text Node
\draw (314,79.4) node [anchor=north west][inner sep=0.75pt] {$-$};
% Text Node
\draw (232,66.4) node [anchor=north west][inner sep=0.75pt] {$-$};
% Text Node
\draw (226,76.4) node [anchor=north west][inner sep=0.75pt] {$-$};
% Text Node
\draw (240,57.4) node [anchor=north west][inner sep=0.75pt] {$-$};
% Text Node
\draw (252,51.4) node [anchor=north west][inner sep=0.75pt] {$-$};
% Text Node
\draw (222,139.4) node [anchor=north west][inner sep=0.75pt] {$+$};
% Text Node
\draw (231,153.4) node [anchor=north west][inner sep=0.75pt] {$+$};
% Text Node
\draw (241,163.4) node [anchor=north west][inner sep=0.75pt] {$+$};
% Text Node
\draw (253,168.4) node [anchor=north west][inner sep=0.75pt] {$+$};
% Text Node
\draw (284.32,167.86) node [anchor=north west][inner sep=0.75pt]
      [rotate=-359.43] {$+$};
% Text Node
\draw (298.37,159.96) node [anchor=north west][inner sep=0.75pt]
      [rotate=-359.37] {$+$};
% Text Node
\draw (309.32,148.14) node [anchor=north west][inner sep=0.75pt]
      [rotate=-358.41] {$+$};
% Text Node
\draw (316.41,137.18) node [anchor=north west][inner sep=0.75pt]
      [rotate=-358.48] {$+$};
% Text Node
\draw (268,169.4) node [anchor=north west][inner sep=0.75pt] {$+$};
% Text Node
\draw (273,119.4) node [anchor=north west][inner sep=0.75pt]
      [font=\Large] [color=white] {${\mathbf{\zeta }}$};

\end{tikzpicture}
  \caption{Esfera metálica cuyos electrones han sido desplazados en la
    dirección de $\bm{\zeta}$ (flecha negra), dejando una ausencia de
    carga en la dirección opuesta induciendo polarización en el
    sistema $\bm P$ (flechas verdes) y una carga
    superficial la cual produce un campo eléctrico
    $\bm{E}(\bm{r},t)$ (flechas azules).}
\label{Polariton}

\end{figure}

En lugar de un sistema semiinfinito, consideremos ahora un sistema
finito consistente en una partícula metálica separada del vacío por
una superficie esférica. Supongamos que perturbamos esta esfera
moviendo todos sus electrones una separación $\bm{\zeta}$ respecto a
su posición de equilibrio, lo cual induce una polarización
$\bm{P}=-n_0 e\bm \zeta$, como ilustra la fig. \ref{Polariton},
El desplazamiento de los electrones hacia uno de los hemisferios de la
esfera genera un exceso de carga negativa en su superficie y un exceso
de carga positiva en el otro, descrita por la densidad de carga
superficial $\sigma=\bm{P}\cdot\hat n$, donde $\hat n$ es un
vector unitario radial. Estas cargas producen un campo eléctrico
$\bm{E}(\bm{r},t) = -(4\pi/3)\bm{P} $ donde empleamos el
{\em factor de despolarización} $-4\pi/3$ de una esfera. Este campo
acelera las cargas de acuerdo a
$md^{2}\bm\zeta/d^{2}t=-e\bm{E}(\bm{r},t)$. Escribiendo al
campo en términos de $\bm\zeta$,
$\bm{E}= -(4\pi n_0 e^{2}/3m)\bm{\zeta}$, la ecuación de movimiento
para $\bm{\zeta}$ se convierte en,
\begin{equation}
  \frac{d^{2}\bm{\zeta}}{d^{2}t}= -\frac{4\pi n_0 e^{2}}{3m}\bm{\zeta}
  \label{polaritondifec}
\end{equation}
de la cual obtenemos la frecuencia $\omega_d$ de las oscilaciones de carga en una
esfera, a las que se les denominan como {\em plasmón dipolar},
\begin{equation}
  \omega_{d}^{2} = \frac{4\pi n e^{2}}{3m} = \frac{\omega^2_{p}}{3}.
\end{equation}

Estos ejemplos muestran que en un metal los electrones pueden animarse
de movimientos colectivos asociados a ciertas frecuencias de
resonancia, las cuales a su vez dependen de la geometría, como
ilustramos estudiando el caso de un sistema infinito, uno semiinfinito
y una esfera. Más aún, si colocamos las partículas metálicas en el
seno de una matriz dieléctrica habría un corrimiento adicional en su
frecuencia de resonancia debido a las cargas inducidas en la
superficie del dieléctrico. Si además hubiese un gran número de
esferas, sus interacciones mutuas a través de los campos
electromagnéticos inducidos podrían generar corrimientos adicionales
de las resonancias.

\begin{figure}
    \centering
    \includegraphics[width = 0.6\textwidth]{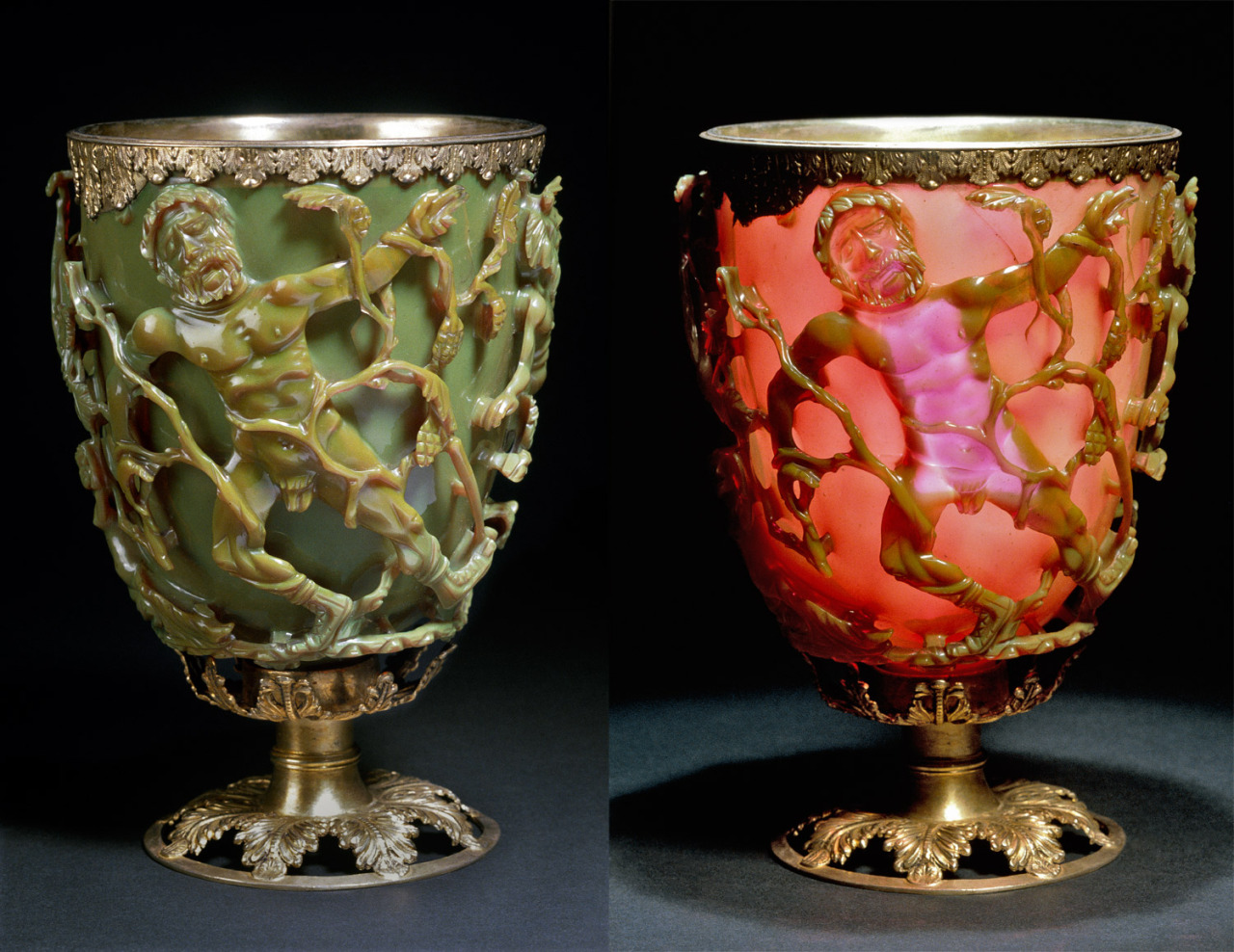}
    \caption{Copa de Licurgo, fabricada en la Roma tardía del siglo IV
      D.C., cuya fabricación resultante muestra propiedades ópticas
      excepcionales.  Es una taza para beber de vidrio que se ve verde
      o roja, dependiendo de cómo es iluminada.  ©The Trustees of the
      British Museum.}
    \label{Lycurgus}
\end{figure}

Un ejemplo excepcional de las propiedades que emergen al generar
\textit{metamateriales} es la copa de Licurgo, una copa de cristal
tallada en la época romana tardía, decorada con un friso que muestra
escenas del mito del Rey Licurgo. La copa que data del siglo IV, D.C.,
se produjo a partir de una pieza en bruto de vidrio soplado de unos
15mm de espesor. Las figuras se cortaron, rectificaron y  unieron a
la pared del recipiente mediante pequeños puentes de vidrio. Aparte
del trabajo artístico realizado en la decoración, la copa es de gran
interés por las propiedades ópticas que muestra. El vidrio se ve de
un color rojo-vino profundo cuando la luz lo atraviesa y de un color verde opaco
cuando la luz que llega a nuestros ojos es reflejada por su
superficie, como muestra la fig. \ref{Lycurgus}. A este fenómeno se le
denomina \textit{dicroísmo}, y de los artefactos de vidrio romano
encontrados, la copa es la que muestra dicho efecto más
intensamente. \cite{LycurgusInvestigation}
Estudios de la composición del vidrio muestran que tales propiedades
son causadas por la presencia de finas partículas de oro,
probablemente en una aleación con plata, dispersadas. Con estudios de
microscopía de transmisión de electrones, TEM, por sus siglas en
inglés, se pudieron determinar tamaños de las partículas de
$\approx 10 nm $.  Se ha encontrado que contiene además partículas de
diferentes metales y de materiales no metálicos. El color se debe al
espectro de reflexión y de transmisión del medio compuesto formado por
vidrio y por las partículas metálicas. Aunque el oro es amarillo, las
partículas de oro embebidas en una matriz de vidrio e interaccionando
entre sí producen un color rojo.

\subsection{Otras geometrías}
\begin{figure}
  \centering
  \includegraphics[width=0.4\textwidth]{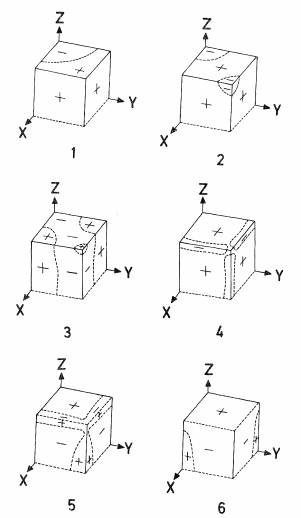}
  \includegraphics[width=0.5\textwidth]{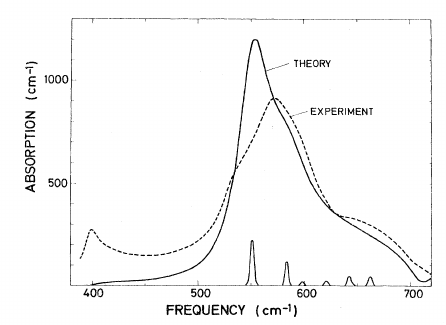}
  \caption{Modos resonantes de un cristal de sal y espectro de
    absorción teórico y experimental.  Los
    signos del lado izquierdo indican la polarización en una de las
    ocho esquinas del cubo de sal correspondiente para cada uno de los
    seis modos principales. La posición y peso de cada resonancia se
    ilustran con las pequeñas gaussianas en la base del panel
    derecho, las cuales generan el espectro teórico al escalarse,
    ensancharse y sumarse.}
  \label{fig:Fuchs}
\end{figure}

Si consideráramos partículas con otras geometrías habría otras
resonancias asociadas a la excitación de modos con varios patrones de
distribución de carga. Por ejemplo, en la figura \ref{fig:Fuchs}
mostramos resultados experimentales y los primeros resultados teóricos
para los modos electromagnéticos esperados en pequeños cubos de sal
\cite{Fuchs}, sus frecuencias de resonancia y su distribución asociada
de carga superficial.
En este caso se encontraron en lugar de un modo dipolar, como vimos
para el caso de la esfera, seis modos principales y unos modos
adicionales con poca fuerza de oscilador, con una polarización cuya
distribución espacial muestra bastante riqueza.

\subsection{Cristales fotónicos}
\begin{figure}
    \begin{subfigure}[b]{0.3\textwidth}
    \centering
    \includegraphics[width=\textwidth]{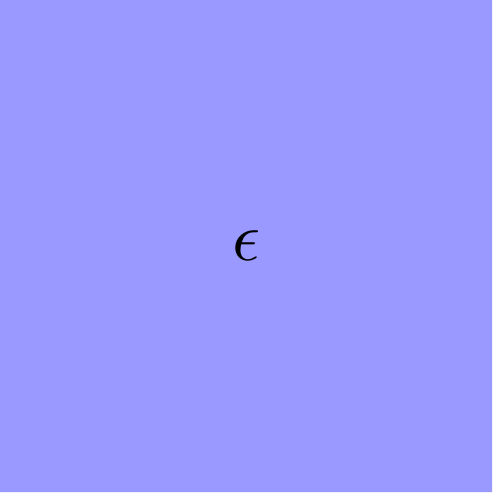}
    \caption{}
    \end{subfigure}
    \hfill
    \begin{subfigure}[b]{0.3\textwidth}
    \centering
    \includegraphics[width=\textwidth]{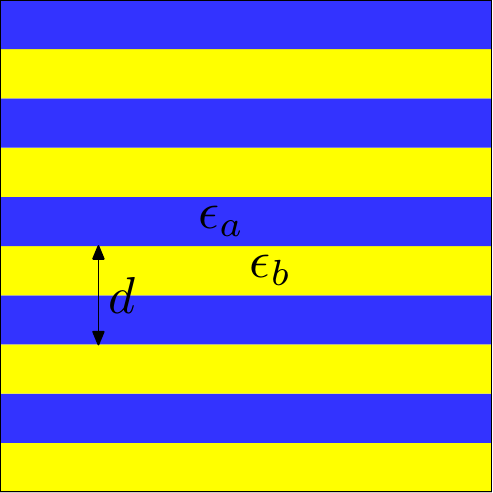}
    \caption{}
    \end{subfigure}
    \hfill
    \begin{subfigure}[b]{0.3\textwidth}
    \centering
    \includegraphics[width=\textwidth]{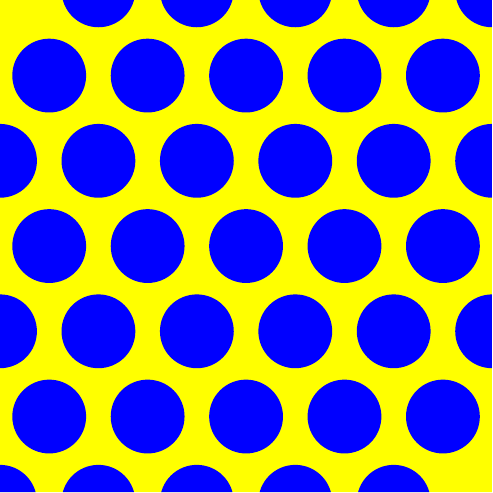}
    \caption{}
    \end{subfigure}
    \hfill
  \newline
  \begin{subfigure}[b]{0.3\textwidth}
    \centering
    \includegraphics[width=\textwidth]{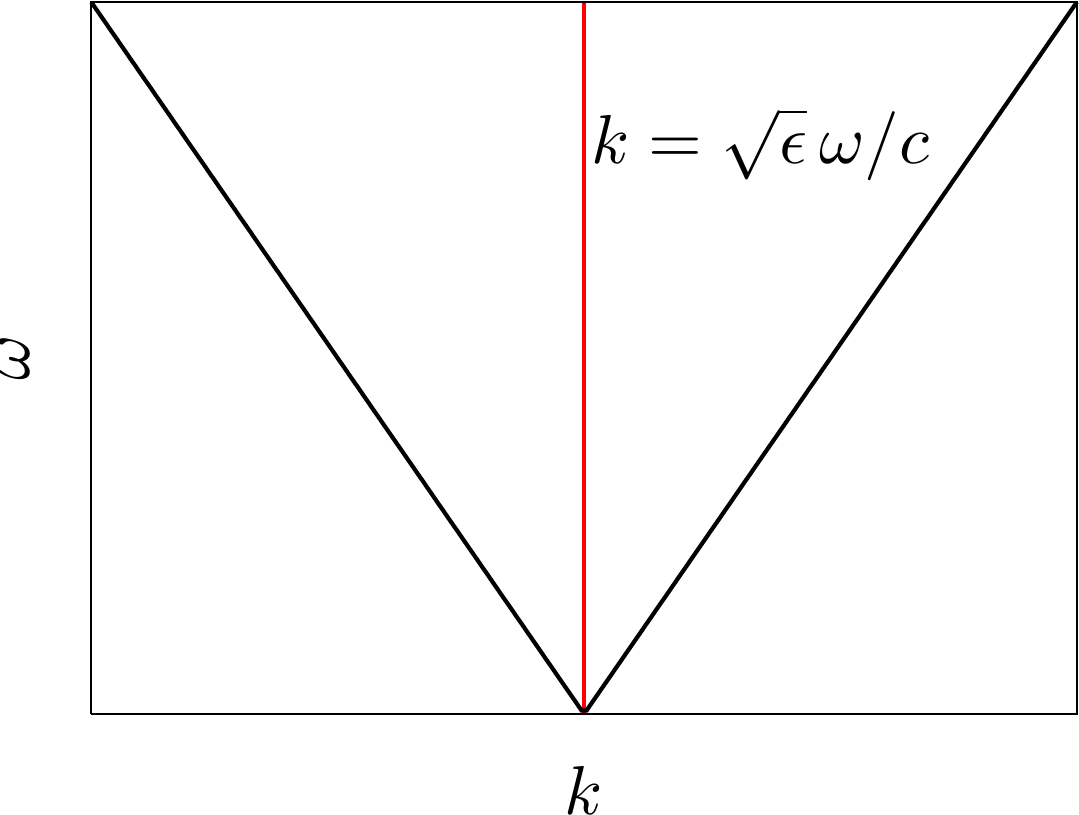}
    \caption{}
    \end{subfigure}
    \hfill
    \begin{subfigure}[b]{0.3\textwidth}
    \centering
    \includegraphics[width=\textwidth]{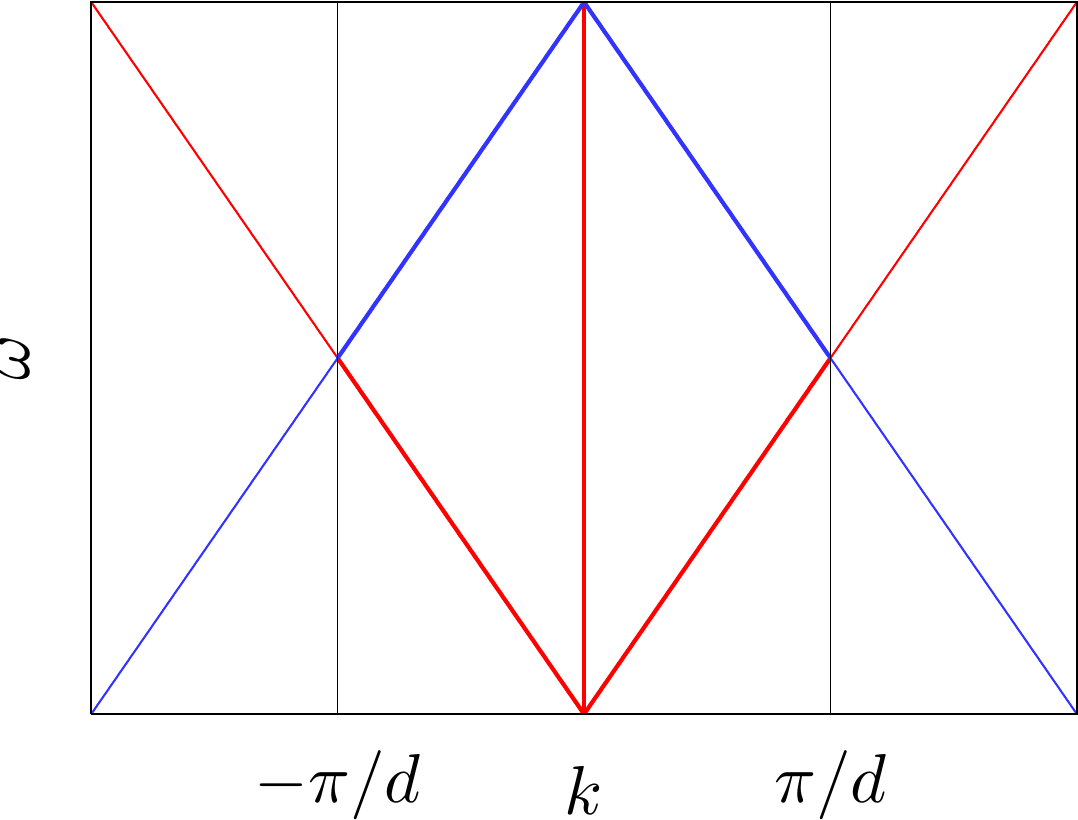}
    \caption{}
    \end{subfigure}
    \hfill
    \begin{subfigure}[b]{0.3\textwidth}
    \centering
    \includegraphics[width=\textwidth]{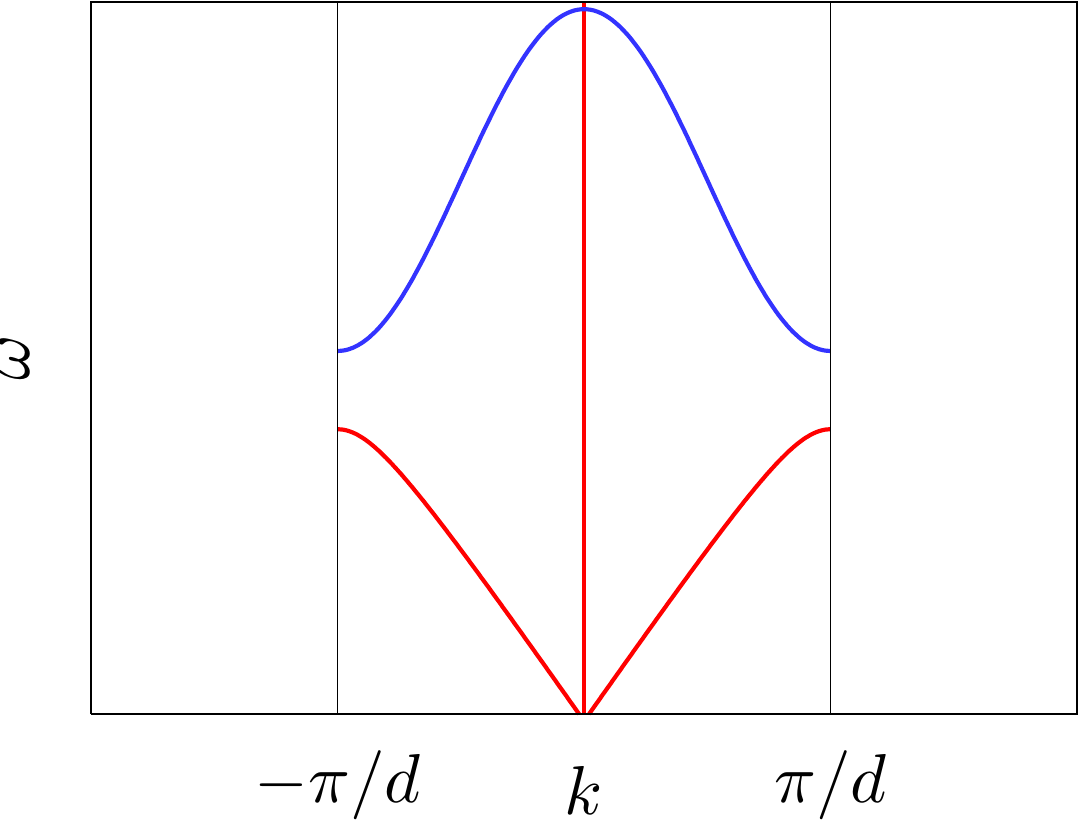}
    \caption{}
    \end{subfigure}
    \hfill
  \caption{(a)Medio dieléctrico homogéneo con respuesta $\epsilon$,
    (b) cristal fotónico unidimensional con funciones respuesta
    $\epsilon_a$ y $\epsilon_b$ y periodo $d$, y (c) cristal fotónico
    con periodicidad en más dimensiones. (d)Relación de dispersión
    $\omega$ vs $k$ de la luz del medio homogéneo. (e)Relaciones de
    dispersión trasladadas por $0, \pm 2\pi/d$ en el espacio
    recíproco mostrando cruzamientos en el borde de la zona de
    Brillouin $k=\pm \pi/d$. (f)El acoplamiento abre una brecha en los
    puntos de degeneración evitando el cruce y formando brechas
    prohibidas.}
  \label{fig:fotonico}
\end{figure}

Consideremos ahora un dieléctrico transparente no dispersivo
homogéneo, como en la figura \ref{fig:fotonico}.
La relación de dispersión de la luz en este medio está dada por
$k^2= \epsilon \omega^2/c^2$ que corresponde a las dos rectas
mostradas en la (fig. \ref{fig:fotonico}d). Si en vez de un
dieléctrico homogéneo tuviéramos un cristal artificial formado por
películas de dos materiales alternados (fig. \ref{fig:fotonico}b), el
ímpetu y el vector de onda ya no serían cantidades conservadas. Las
reflexiones múltiples en las interfaces producirían ondas esparcidas
en que el vector de onda cambiaría $k\to k+2\pi n/d$ para enteros
positivos y negativos $n$ (fig. \ref{fig:fotonico}e), dando lugar a
puntos de degeneración en que se cruzan las distintas réplicas de la
relación de dispersión. El acoplamiento entre los campos esparcidos
rompe la degeneración y abre {\em brechas fotónicas} evitando los
cruces y dando origen a una relación de dispersión
(fig. \ref{fig:fotonico}f) organizada en {\em bandas fotónicas}
análogas a las bandas electrónicas que describen la propagación de
electrones en sólidos cristalinos. Algo similar sucedería si la
periodicidad fuese bidimensional o tridimensional
(fig. \ref{fig:fotonico}c) en cuyo caso podrían producirse brechas
omnidireccionales en las que la luz no se propaga en ninguna
dirección. Las brechas fotónicas explican algunos fenómenos naturales,
como la iridiscencia en los caparazones de diversos insectos y los
colores de las alas de las mariposas, colores producidos no por
pigmentos que absorben la luz, sino por pequeñas estructuras
dieléctricas transparentes que forman cristales fotónicos con regiones
de frecuencia en que la luz es fuertemente reflejada por corresponder
a brechas en que no se puede propagar. Estos colores se llaman por su
origen {\em colores estructurales}. Introduciendo {\em defectos} en
cristales fotónicos se pueden generar sitios en que la luz puede ser
atrapada, hecho que ha encontrado aplicaciones tales como la
elaboración de fibras ópticas fotónicas.

\subsection{Materiales izquierdos}
\begin{figure}
  \centering
  \includegraphics{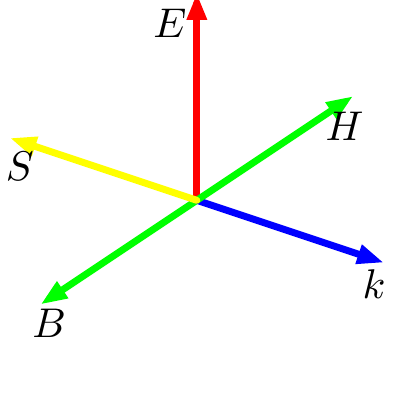}
  \caption{Relación entre las direcciones del campo eléctrico $\bm E$,
    densidad de flujo magnético $\bm B$, vector de onda $\bm k$, campo
    magnético $\bm H$ y vector de Poynting $\bm S$ en un metamaterial
    izquierdo isotrópico. Como $(\bm E, \bm B, \bm k)$ y
    $(\bm E, \bm H,\bm S)$ son triadas ordenadas derechas, y $\bm B$ y
    $\bm H$ son antiparalelos, entonces $\bm k$ y $\bm S$ son
    antiparalelos.  }
  \label{fig:triada}
\end{figure}

\begin{figure}
  \centering
  \includegraphics{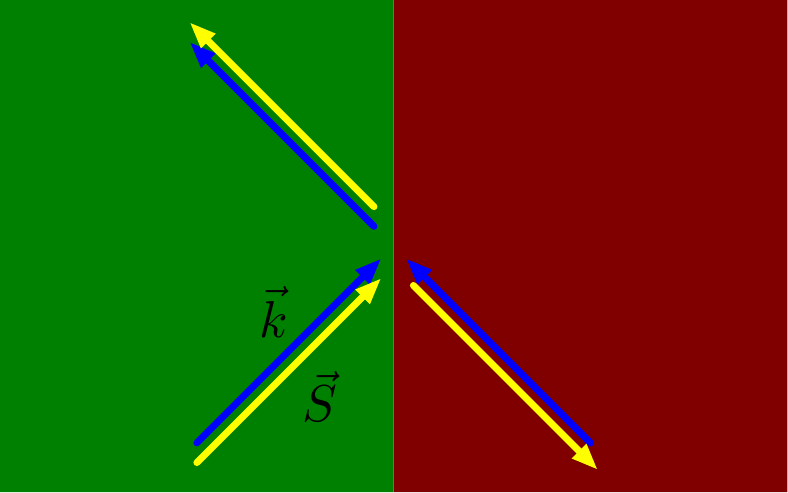}
  \caption{Una onda incide desde un medio normal (verde) sobre la
    interfaz que lo separa de un medio con permitividad y
    permeabilidad negativas (rojo). Se muestran esquemáticamente las
    direcciones de los vectores de onda (flechas azules) y de los
    vectores de Poynting (flechas amarillas) de las ondas incidente,
    reflejada y transmitida. La proyección sobre la interfaz de los
    tres vectores de onda debe coincidir. El flujo de energía de la
    onda transmitida debe alejarse de la interfaz. Por ello, la
    componente normal del vector de onda transmitido apunta hacia la
    interfaz.}
  \label{fig:refneg}
\end{figure}

\begin{figure}
  \centering
  \includegraphics{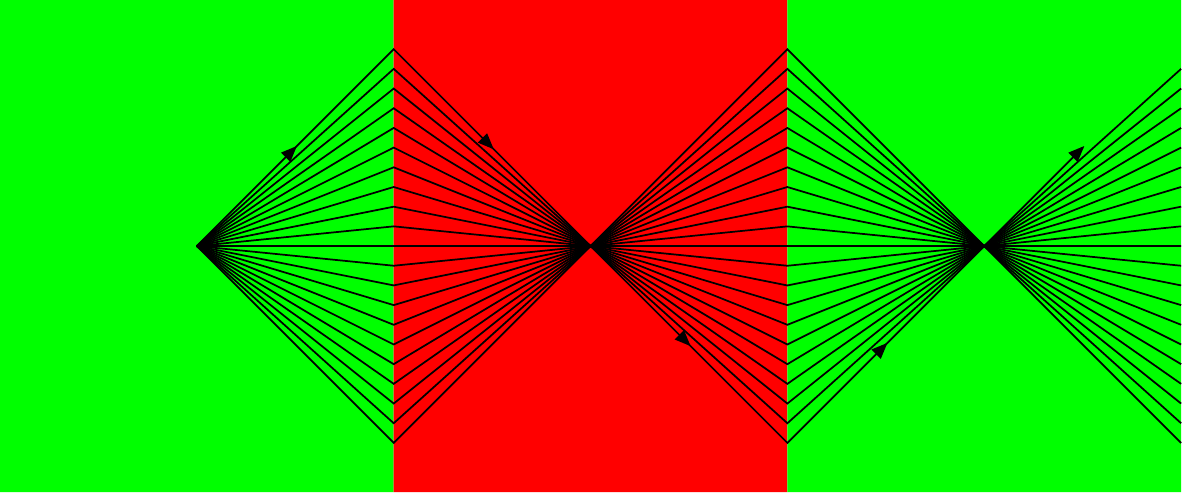}
  \caption{Lente consistente en una película plana de un metamaterial
    izquierdo (rojo) en el seno de un material ordinario (verde). Se
    ilustra la trayectoria de varios rayos que emergen de una fuente
    puntual de luz y convergen en un punto imagen. }
  \label{fig:lenteplana}
\end{figure}

Consideremos ahora un material cuya permitividad $\epsilon(\omega)$
dependa de la frecuencia y tenga un comportamiento resonante. La
relación de dispersión $k^2=\epsilon\omega^2/c^2$ implica que al pasar
la resonancia, cuando $\epsilon$ adquiere valores negativos, $k$ se
vuelve imaginario y la luz no puede propagarse. Esto explica la
aparición del color en los materiales comunes, en que hay frecuencias
características de cada material en que absorben luz y justo arriba
hay frecuencias en que no se puede propagar. Si el material tuviera
además una respuesta magnética $\mu\ne 1$ la relación de dispersión
cambiaría a $k^2=\epsilon\mu\omega^2/c^2$. Si tanto $\epsilon$ como
$\mu$ tuvieran resonancias cercanas, arriba de éstas podría suceder
que ambas fueran negativas, $\epsilon<0$ y $\mu<0$. En este caso, su
producto $\epsilon\mu>0$ sería positivo y sí podría haber propagación
con un vector de onda real. Sin embargo, esta propagación sería
curiosa. A partir de las ecuaciones de Maxwell, por ejemplo,
de las leyes de Faraday y de Gauss magnética, sabemos que para una onda plana,
el campo eléctrico $\bm E$, la densidad de flujo magnético $\bm B$ y
el vector de onda $\bm k$ forman una triada ordenada derecha, como
ilustra la figura \ref{fig:triada}.
Sin embargo, la definición del vector de Poynting y la ley de
Ampère-Maxwell implican que el campo eléctrico $\bm E$, el campo
magnético $\bm H$ y el flujo de energía $\bm S$ también forman una
triada ordenada derecha. Sin embargo, si $\mu<0$, entonces $\bm B$ y
$\bm H$ apuntan en direcciones opuestas. Luego, $\bm S$ ¡apunta en la
dirección opuesta a $\bm k$! La dirección en que avanza la fase de la
onda es opuesta a la dirección en que avanza la energía. Esto sólo
puede ser posible si la velocidad de grupo es opuesta a la velocidad
de fase. Una consecuencia curiosa de este resultado se manifiesta
cuando una onda se refracta en una superficie plana. La ley de
conservación del ímpetu asociada a una simetría translacional implica
que las proyecciones del vector de onda $\bm k_\|$ a lo largo de la
superficie deben coincidir para las ondas reflejada,  transmitida e
incidente. De aquí se derivan las leyes de la reflexión y de
Snell. Sin embargo, la {\em causalidad} requiere que las ondas
esparcidas por la superficie, la onda incidente y la onda reflejada,
deben tener un flujo de energía que se aleja de la superficie. Ello
implica que cuando incide luz desde un medio ordinario hacia un medio
con $\epsilon<0$ y $\mu<0$, la componente normal del vector de onda de
la onda transmitida ¡debe apuntar hacia la superficie!, como ilustra
la fig. \ref{fig:refneg}.
De esta figura podemos inferir que una onda que incide viajando hacia
arriba se refracta hacia abajo y viceversa. Eso lleva a plantear
dispositivos como el ilustrado en la fig. \ref{fig:lenteplana},
consistente en una película plana de un metamaterial izquierdo con
$\epsilon<0$ y $\mu<0$.
La luz que emerge de una fuente puntual y viaja hacia la derecha y
hacia arriba se refracta hacia abajo mientras que luz que viaja hacia
abajo se refracta hacia arriba. Algo análogo sucede al emerger de la
película. Es posible entonces que todos los rayos que parten de la
fuente luminosa converjan en un punto, la imagen de la fuente formada
por una lente plana.

\begin{figure}
  \centering
  \includegraphics[width=.7\textwidth]{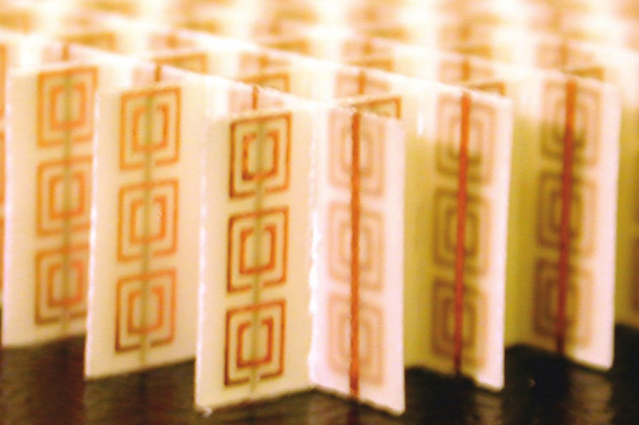}
  \caption{Metamaterial izquierdo formado por una red de parejas de
    anillos conductores interrumpidos ({\em split rings}) y pistas
    conductoras rectas sobre un dieléctrico. (Tomada de la ref. \cite{shelby})}
  \label{fig:rings}
\end{figure}

Desafortunadamente, no existen materiales naturales en los que tanto
la permitividad como la permeabilidad sean negativas a la misma
frecuencia. Sin embargo, hay metamateriales artificiales que pueden
describirse por una permitividad y permeabilidad efectiva que sí
cumplan esta condición. La figura \ref{fig:rings} muestra un ejemplo
formado por un arreglo de parejas de anillos interrumpidos que
funcionan como un circuito LC resonante. La corriente recorriendo los
anillos produce un dipolo magnético, y debido a su interrupción,
produce una acumulación de cargas que lo acopla con el otro
anillo. Este circuito tiene una resonancia arriba de la cual la
permeabilidad macroscópica es negativa. Por otro lado, una serie de
pistas rectas permiten que el material se comporte en la dirección
vertical como un conductor, por lo cual la permitividad es negativa
abajo de la frecuencia de plasma efectiva.
\subsection{Partículas dieléctricas}
Hemos visto arriba que partículas metálicas pequeñas pueden tener
resonancias plasmónicas cuyas frecuencias dependen en general de su
composición y de su geometría. También partículas dieléctricas pueden
tener resonancias aunque estén formadas por materiales no dispersivos,
siempre y cuando la longitud de onda de la luz en su interior sea
conmensurable con su tamaño. Estas resonancias se deben a la
interferencia constructiva entre ondas múltiplemente reflejadas por
sus superficies. Por ejemplo, partículas esféricas o cilíndricas
muestran {\em resonancias de Mie} cuando el perímetro de su sección
transversal es cercano a un múltiplo de la longitud de onda.  Una
ventaja de estas resonancias sobre las resonancias plasmónicas para
diseñar y construir dispositivos fotónicos es que las pérdidas de
energía debidas a la absorción dentro del material son menores que las
pérdidas óhmicas que suelen mostrar los metales. Sin embargo, estas
resonancias requieren que las partículas tengan un tamaño
relativamente grande conmensurable con la longitud de onda en su
interior. Sin embargo, si se emplean materiales con un índice de
refracción alto, la longitud de onda dentro de estos materiales puede
ser mucho menor que la correspondiente al espacio vacío, permitiendo
así resonancias dieléctricas en partículas de tamaño muy pequeñas, en
analogía a las resonancias plasmónicas.
\subsection{Metasuperficies}
Las funciones respuesta de una partícula cambian de signo conforme la
frecuencia de la luz pasa de ser menor a ser mayor a su frecuencia de
resonancia. Por tanto, la fase que adquiere un haz luminoso al pasar a
través de una superficie cubierta por partículas depende muy
sensiblemente de la cercanía de la frecuencia a la frecuencia de
resonancia de las partículas, la cual a su vez, depende de la
geometría. Por tanto, modulando la geometría de las partículas a lo
largo de la superficie, puede modularse la fase que adquiere la luz en
forma análoga a como el ancho variable de una lente o de un prisma
modula la fase de los rayos de luz que los atraviesan.

La ley de Snell usual implica que a lo largo de una interfaz uniforme
hay un empatamiento de fases $\phi^\alpha(\bm r_\|)=\bm k^\alpha_\|\cdot\bm
r_\|$ entre la onda incidente, la onda reflejada y la
onda transmitida, por lo cual los vectores de onda $\bm k^\alpha_\|$
proyectados sobre la superficie son iguales para las tres ondas $\alpha=i,r,t$. Sin
embargo, si la superficie no es uniforme y a lo largo de ésta la onda
transmitida y/o reflejada adquiere una fase adicional $\psi^\alpha(\bm
r_\|)$ a la de la onda incidente, $\phi^\alpha(\bm
r_\|)=\bm k^i_\|\cdot\bm r_\|+\psi^\alpha(\bm r_\|)\approx (\bm
k^i_\|+\nabla_\|\psi^\alpha(0))\cdot\bm r_\|$ la ley de Snell debe
generalizarse,
\begin{equation}
  \label{eq:snell}
  \bm k^\alpha_\|=\bm k^i_\|+\nabla_\|\psi^\alpha,\quad(\alpha=r,t)
\end{equation}
i.e., el ímpetu paralelo a la interfaz adquiere una contribución
debida a la variación de la fase adicional $\psi^\alpha$. Por lo
tanto, modulando la fase de una onda a lo largo de una superficie
podemos manipular la dirección de la luz transmitida o reflejada. Para
esto se pueden colocar partículas con un índice de refracción grande
sobre una superficie ordinaria y modificar a lo largo de esta su
geometría, orientación o densidad, dando lugar a una {\em metasuperficie}.

\begin{figure}
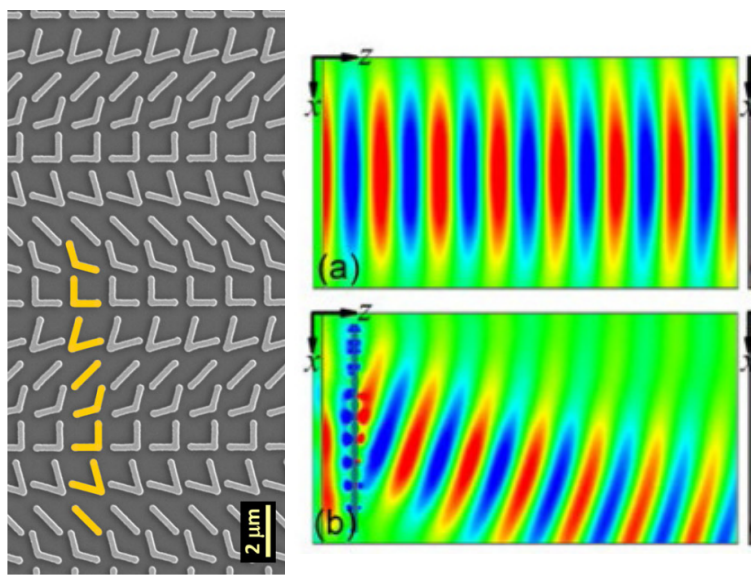

  \centering
  \includegraphics[width=0.5\textwidth,angle=90,valign=c]{fig12a}
  \includegraphics[width=0.4\textwidth,valign=c]{fig12b}
  \caption{Metasuperficie formada por un arreglo de partículas en
    forma de L con ángulos variables (izquierda), capaz de separar un
    haz luminoso en sus componentes con distintas polarizaciones
    (derecha).(Tomada de la ref. \cite{metasup}). }
  \label{fig:metasup}
\end{figure}

\begin{figure}
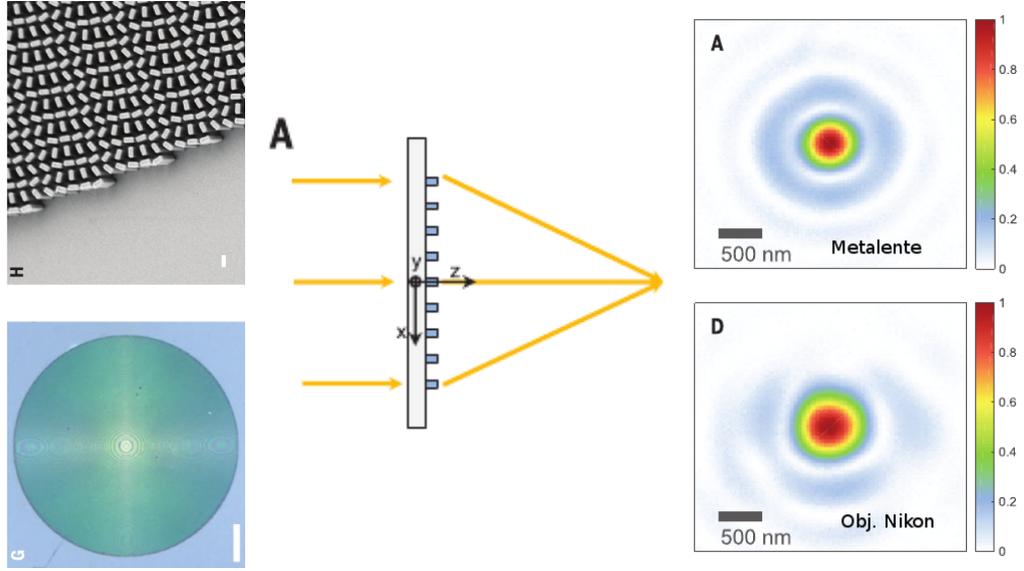

  \centering
  \includegraphics[width=.5\textwidth,angle=90, valign=c]{fig13a}
  \includegraphics[height=.3\textwidth,valign=c]{fig13b}
  \includegraphics[height=.5\textwidth,valign=c]{fig13c}
\caption{Metalente formada por un arreglo de prismas de alto índice
    de refracción con orientaciones variables (izquierda). Refracción
    de la luz debido a la modulación de fase al atravesar el metalente
    (centro). Imagen de una
    fuente puntual formada por la metalente y por una lente
    convencional (derecha). (Tomada de la ref. \cite{metalente})}
  \label{fig:metalente}
\end{figure}

Como las resonancias dependen también de la polarización de la luz,
este efecto puede usarse para desviar haces de luz de acuerdo a su
polarización. En la fig. \ref{fig:metasup} mostramos una {\em
  metasuperficie} formada por partículas en forma de $L$ cuya
respuesta difiere cuando es iluminada con polarización horizontal o
vertical, y que por lo tanto puede
separar un haz de luz no polarizada en dos haces con polarizaciones
perpendiculares.
Otros sistemas pueden separar la luz de acuerdo a su helicidad, en
haces con polarización circular derecha o circular izquierda.
Mediante otros arreglos se puede variar la fase a lo largo de la
dirección radial, de forma de hacer converger rayos que arriben en la
dirección normal sobre un punto, su foco, creando así una metalente,
como la que ilustra la figura \ref{fig:metalente}.

\section{Teoría}

Las propiedades ópticas de materiales compuestos como los presentados
arriba, metamateriales, cristales fotónicos, materiales izquierdos,
etc., con propiedades en ocasiones exóticas, están determinadas no
sólo por su composición, sino también por su geometría. Propiedades
como las relaciones de dispersión de los modos electromagnéticos que
se propagan a través de un metamaterial extendido, las amplitudes de
reflexión y transmisión, y relaciones de dispersión de modos
electromagnéticos confinados a la superficie de sistemas con
fronteras, las secciones transversales de dispersión, absorción y
extinción de partículas formadas por partículas, pueden ser expresadas
en términos del operador dieléctrico macroscópico del compuesto a
través de las soluciones de las ecuaciones de Maxwell en dicho
material. En esta sección presentaremos un formalismo para obtener la
respuesta macroscópica en términos de la respuesta microscópica.

\subsection{Proyectores promedio y fluctuación}
Primero notamos que el campo electromagnético dentro de un material inhomogéneo
tiene oscilaciones relacionadas con su {\em textura}  y la
escala de variación espacial de estas oscilaciones es del orden del
tamaño de las partículas que forman el compuesto así como de las
distancias entre partículas vecinas.  Llamamos campo macroscópico a
aquel del cual hemos eliminado dichas fluctuaciones. El campo
macroscópico puede tener oscilaciones espaciales, pero éstas están
asociadas a las oscilaciones temporales del campo y a la longitud de
onda finita de los campos que se propagan. En todo caso, es
conveniente introducir dos operadores, el {\em promedio}
$\hat {\mathcal P}_p$ y la {\em fluctuación} $\hat {\mathcal P}_f$,
tales que al actuar sobre un campo arbitrario $\bm F$ producen el
campo promedio $\bm F_p=\hat {\mathcal P}_p \bm F$ y el campo
fluctuante $\bm F_f=\hat {\mathcal P}_f \bm F$. Existen muchas formas
de definir promedio. En sistemas desordenados podríamos usar el
promedio de ensamble, es decir, sumar sobre $N$ realizaciones del
sistema y dividir entre $N$, tomando el límite $N\to\infty$. En
sistemas dinámicos como un fluido, y para campos que oscilen
lentamente con respecto a los tiempos característicos en que cambia el
sistema, podríamos emplear un promedio temporal. Para otro tipo de
sistemas podríamos tomar un promedio espacial o aplicar un filtro
pasabajo en el espacio recíproco. Lo que debe ser claro es que un
campo se promedia cuando se le remueven las fluctuaciones, i.e.,
$\hat{\mathcal P}_p=\hat {\bm 1}-\hat{\mathcal P}_f$, con
$\hat {\bm 1}$ el operador identidad. Si pretendiéramos remover las
fluctuaciones de un campo que ya hemos promediado, encontraríamos
que no queda nada por remover. Esto implica que el operador promedio
es idempotente $\hat{\mathcal P}_p^2=\hat{\mathcal P}_p$, i.e., el
promedio del promedio es el promedio. Análogamente, las fluctuaciones
son lo que queda al remover el promedio. Por lo tanto, el operador
fluctuación también es idempotente,
$\hat{\mathcal P}_f^2=\hat{\mathcal P}_f$. Finalmente, si eliminamos las
fluctuaciones y el promedio, no nos queda nada,
$\hat{\mathcal P}_p\hat{\mathcal P}_f=0$,
$\hat{\mathcal P}_f\hat{\mathcal P}_p=0$.

Los resultados previos muestran que $\hat{\mathcal P}_p$ y
$\hat{\mathcal P}_f$ son proyectores que mandan a un campo al
subespacio de los campos promedio y al subespacio de los campos
fluctuantes respectivamente, y que el espacio donde vive originalmente
el campo vectorial es una suma directa de estos dos subespacios. Esto
permite escribir formalmente a un campo arbitrario como si fuera un
vector de dos componentes,
\begin{equation}
  \label{eq:vectorpf}
  \bm F=
  \begin{pmatrix}
    \bm F_p\\\bm F_f
  \end{pmatrix},
\end{equation}
aunque cada componente en sí no es un número sino un campo
vectorial. Análogamente, las funciones respuesta pueden representarse
como operadores lineales en términos de matrices de dos por dos. Así,
la ecuación $\bm D=\hat{\bm\epsilon} \bm E$ puede escribirse como una
ecuación matricial
\begin{equation}
  \label{eq:DeE}
  \begin{pmatrix}
    \bm D_p\\\bm D_f
  \end{pmatrix}
  =
  \begin{pmatrix}
    \hat{\bm  \epsilon}_{pp}&\hat{\bm  \epsilon}_{pf}\\\hat{\bm
      \epsilon}_{fp}&\hat{\bm  \epsilon}_{ff}
  \end{pmatrix}
  \begin{pmatrix}
    \bm E_p\\\bm E_f
  \end{pmatrix}.
\end{equation}
Aquí, hemos definido $\hat{\bm O}_{\alpha\beta}=\hat{\mathcal
  P}_\alpha\hat {\bm O} \hat{\mathcal P}_\beta$ ($\alpha,\beta=p,f$)
para cualquier operador $\hat{\bm O}$.
Interpretaremos a la ec. \eqref{eq:DeE} como un ecuación material {\em
  microscópica}, pues incorpora las fluctuaciones espaciales derivadas
de la textura del material. La correspondiente ecuación macroscópica
sería
\begin{equation}
  \label{eq:DeEmacro}
  \bm D_M=\hat{\bm \epsilon}_M\bm E_M,
\end{equation}
donde identificamos a los campos macroscópicos como los campos
promediados y por lo tanto, libres de fluctuaciones, $\bm D_M\equiv \bm D_p$,
$\bm E_M\equiv\bm E_p$. En general, $\hat{\bm\epsilon}_M$ no es el
promedio $\hat{\bm\epsilon}_{pp}$ de $\hat{\bm \epsilon}$, pues puede
haber {\em correlaciones} entre las fluctuaciones espaciales de $\epsilon(\bm r)$
y del campo eléctrico $\bm E(\bm r)$.

% cambié a, por lo tanto, por que en este caso, por lo que antes se ha
% dicho los campos promediados son libres de fluctuaciones.

\subsection{Proyectores longitudinal y transversal}
Por otro lado, recordemos que de acuerdo al teorema de Helmholtz, todo
campo vectorial $\bm F=\bm F^L+\bm F^T$ puede escribirse como la suma
de dos contribuciones, un campo longitudinal
$\bm F^L=\hat{\mathcal P^L}\bm F$ y un campo transversal
$\bm F^T=\hat{\mathcal P^T}\bm F$, que cumplen las ecuaciones
\begin{subequations}
  \label{eq:Helmholtz}
  \begin{align}
    \label{eq:Helmholtza}\nabla\times\bm F^L&=0,\\
    \label{eq:Helmholtzb}\nabla\cdot\bm F^L&=\nabla\cdot\bm F,\\
    \label{eq:Helmholtzc}\nabla\times\bm F^T&=\nabla\times\bm F,\\
    \label{eq:Helmholtzd}\nabla\cdot\bm F^T&=0.
  \end{align}
\end{subequations}
Para obtener los proyectores $\hat{\mathcal P}^L$ y  $\hat{\mathcal
  P}^T$ podemos empezar con la ec. \eqref{eq:Helmholtza}, la cual implica
que $\bm F^L$ puede derivarse de algún potencial escalar $\Phi$ como
$\bm F^L=-\nabla\Phi$. Luego, la ecuación \eqref{eq:Helmholtzb}
implica que el potencial obedece la ec. de Poisson,
$\nabla^2\Phi=-\nabla\cdot\bm F$, cuya solución formal es
$\Phi=-\hat\nabla^{-2}\hat\nabla\cdot\bm F$. Aquí, $\hat\nabla^{-2}$
representa el operador inverso al laplaciano, el cual puede escribirse
en el espacio real como un operador integral cuyo kernel en 3D es el
potencial coulombiano $-1/(4\pi|\bm r-\bm r'|)$. Finalmente,
obtenemos $\bm F^L=\hat\nabla\hat\nabla^{-2}\hat\nabla\cdot\bm F$ de
donde identificamos al proyector longitudinal
\begin{equation}
  \label{eq:PL}
  \hat{\mathcal P}^L=\hat\nabla\hat\nabla^{-2}\hat\nabla\cdot .
\end{equation}
Análogamente, podemos identificar al proyector transversal
\begin{equation}
  \label{eq:PT}
  \hat{\mathcal P}^T=\hat{\bm 1}-\hat{\mathcal P}^L=-\hat\nabla\times \hat\nabla^{-2}\hat\nabla\times.
\end{equation}

\subsection{Caso no retardado}\label{sec:noret}
Para obtener la respuesta macroscópica $\hat{\bm \epsilon}_M$, que
como hemos mencionado no es el simple promedio de la respuesta microscópica
$\hat{\bm\epsilon}$, recurrimos a las ecuaciones de Maxwell.
Consideremos un material formado por inclusiones muy pequeñas
y muy cercanas entre sí, cuyo tamaño y separación sean mucho menores que la
longitud de onda característica de la luz $\lambda_0=2\pi c/\omega$ a
una frecuencia $\omega$ dada. En este caso podemos ignorar el {\em
  retardamiento}, tomar el límite $c\to\infty$ en la ecuación de
Faraday y tratar al campo eléctrico como si fuera un campo puramente
{\em longitudinal}, $\bm E=\bm E^L$, derivable de un potencial escalar.
Por tanto, la proyección longitudinal del desplazamiento, $\bm
D^L=\hat{\mathcal P}^L\hat\epsilon \bm E=\hat{\mathcal P}^L\hat\epsilon
\bm E^L=\hat{\mathcal P}^L\hat\epsilon\hat{\mathcal P}^L \bm
E^L$,
\begin{equation}
  \label{eq:DLvsEL}
  \bm D^L=\hat{\bm\epsilon}^{LL}\bm E^L,
\end{equation}
cumple las mismas ecuaciones que el campo eléctrico
longitudinal externo,
\begin{equation}
  \label{eq:DvsEext}
  \begin{split}
    \nabla\times\bm D^L&=0=\nabla\times\bm E_{\text{ex}}^{L},\\
    \nabla\cdot\bm D^L&=4\pi\rho^{\text{ex}}=\nabla\cdot\bm E^{L}_{\text{ex}}.
  \end{split}
\end{equation}
Aquí, hemos definido
$\hat{\bm O}^{\alpha\beta}=\hat{\mathcal P}^\alpha\hat {\bm O}
\hat{\mathcal P}^\beta$ ($\alpha,\beta=L,T$) para cualquier operador
$\hat{\bm O}$. Entonces, podemos identificar a $\bm D^L$ con el
campo eléctrico longitudinal externo. Siendo un campo externo, sus
fuentes son únicamente las cargas externas $\rho^{\text{ex}}$, las
cuales no tienen absolutamente nada que ver con la composición del
material ni con la disposición de sus componentes. En particular,
$\bm D^L$ no tiene fluctuaciones espaciales derivadas de la textura
del material, $\bm D^L=\bm D^L_p= \hat{\mathcal P_p} D^L$. Despejando
el campo eléctrico de la ec. \eqref{eq:DLvsEL} obtenemos
\begin{equation}
  \label{eq:ELvsDL}
  \bm E^L=(\hat{\bm\epsilon}^{LL})^{-1}\bm D^L,
\end{equation}
y promediando ambos lados de la ecuación, usando el hecho de que $\bm
D^L$ no tiene fluctuaciones y que $\hat{\mathcal P}_p$ es idempotente,
obtenemos
\begin{equation}
  \label{eq:ELpvsDLp}
  \bm E^L_p=(\hat{\bm\epsilon}^{LL})^{-1}_{pp}\bm D^L_p.
\end{equation}
Finalmente, identificando los campos promedios con los campos
macroscópicos y su relación con la respuesta macroscópica,
podemos identificar\cite{ElectromagneticResponseofSystemwithSpatialFluctuations}
\begin{equation}
  \label{eq:epsLLM}
  (\hat{\bm \epsilon}_M^{LL})^{-1}=(\hat{\bm\epsilon}^{LL})^{-1}_{pp}.
\end{equation}
Podemos leer este resultado de la siguiente manera: el inverso de la
proyección longitudinal de la respuesta dieléctrica macroscópica es
igual al promedio del inverso de la proyección longitudinal de la
respuesta microscópica.\cite{guilleOE}

\subsection{Caso retardado}\label{sec:ret}
\begin{figure}
  \centering
  \includegraphics[width=0.8\textwidth]{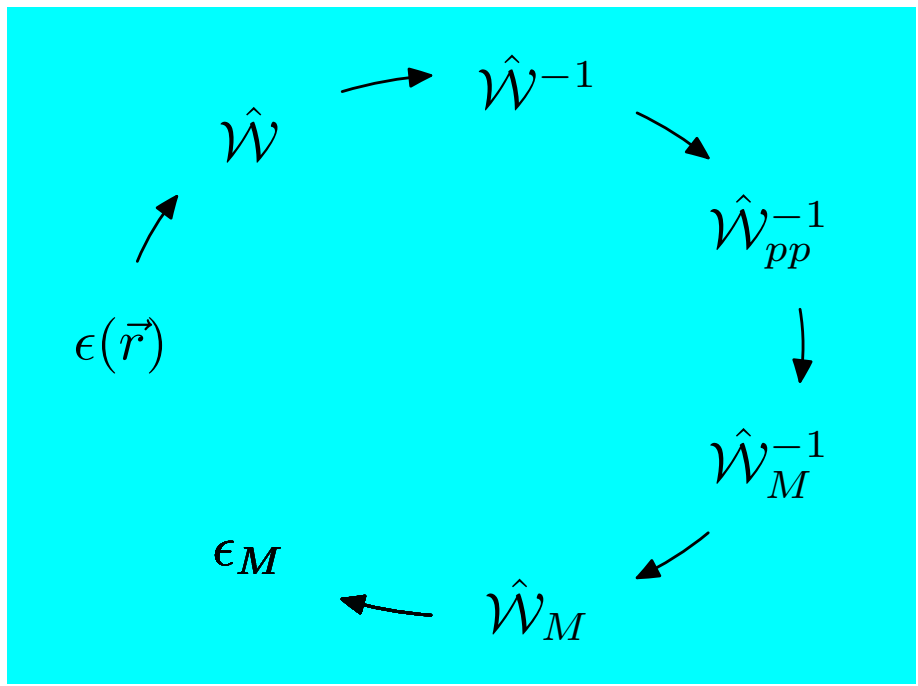}
  \caption{Pasos para obtener la respuesta macroscópica
    $\hat{\bm \epsilon}_M$ a partir de la respuesta microscópica
    $\epsilon(\bm r)$ incluyendo efectos de retardamiento. Primero
    construimos el operador de onda, lo
    invertimos, lo promediamos, identificamos el operador de onda
    macroscópico inverso, lo invertimos y finalmente identificamos el
    operador dieléctrico macroscópico.}
  \label{fig:circulo}
\end{figure}

Como hemos mostrado, para obtener la respuesta
macroscópica de un sistema no basta con promediar cualquier función
respuesta. Por ejemplo, el promedio de $\bm\epsilon$ no tiene
significado como respuesta macroscópica. Sin embargo, si logramos
encontrar un operador que responda a una excitación {\em externa},
la cual no tiene fluctuaciones espaciales asociadas a la textura
microscópica del material, su promedio nos proporciona la
respuesta correspondiente macroscópica. Para dar un ejemplo de éste
proceso, a continuación obtendremos la respuesta macroscópica en el
caso en que no podemos ignorar el retardamiento. Tomando el
rotacional de la ley de Faraday y sustituyendo la ec. de Ampère-Maxwell
podemos obtener una ecuación de onda con fuentes, que podemos escribir
como
\begin{equation}
  \label{eq:onda}
  \hat{\mathcal W}\bm E=\frac{4\pi}{i\omega}\bm j^{\text{ex}},
\end{equation}
donde
\begin{equation}
  \label{eq:opW}
  \hat{\mathcal
    W}=\hat{\bm\epsilon}+\frac{c^2}{\omega^2}\nabla^2\hat{\mathcal P}^T
\end{equation}
es una generalización del {\em operador de onda}.
Podemos resolver la ec. \eqref{eq:onda} formalmente para obtener
el campo en el material
\begin{equation}
  \label{eq:ondasol}
  \bm E=\frac{4\pi}{i\omega}\hat{\mathcal W}^{-1} \bm j^{\text{ex}},
\end{equation}
e interpretar al inverso del operador de onda
$\hat{\mathcal W}^{-1}$ como una respuesta a la excitación externa
$\bm j^{\text{ex}}$, y a su promedio como la
respuesta macroscópica a la corriente externa,
\begin{equation}
  \label{eq:WM}
  \hat{\mathcal W_M}^{-1}=\hat{\mathcal W}^{-1}_{pp}.
\end{equation}
Interpretando al operador de onda macroscópico en términos de la
permitividad macroscópica
\begin{equation}
  \label{eq:opWM}
  \hat{\mathcal
    W}_M=\hat{\bm\epsilon}_M+\frac{c^2}{\omega^2}\nabla^2\hat{\mathcal P}^T
\end{equation}
podemos despejarla tras invertir la ecuación
\eqref{eq:WM}. Formalmente, podemos obtener la permitividad
macroscópica a partir de la permitividad microscópica siguiendo los
pasos indicados en la figura \ref{fig:circulo}.
A partir de la respuesta dieléctrica de las componentes de nuestro
material podemos construir el operador de onda, lo invertimos, lo
promediamos, lo identificamos en términos del operador de onda
macroscópico, lo volvemos a invertir y finalmente obtenemos la
respuesta dieléctrica  macroscópica.\cite{samuel}

\subsection{Sistema binario periódico sin retardamiento}
\label{binario}
Consideremos ahora un sistema periódico hecho de dos materiales,
digamos, de partículas de un material $B$ embebidas en una matriz de
un material $A$. La respuesta dieléctrica {\em microscópica} es en
este caso
\begin{equation}
  \label{eq:epsmic}
  \epsilon(\bm r)=\left\{
      \begin{matrix}
        \epsilon_A\,\quad\text{si $\bm r\in A$}\\
        \epsilon_B\,\quad\text{si $\bm r\in B$}
      \end{matrix}
    \right.
\end{equation}
la cual puede escribirse como
\begin{equation}
  \label{eq:epsmic1}
  \epsilon(\bm r)=\frac{\epsilon_A}{u}(u-B(\bm r)),
\end{equation}
donde
\begin{equation}
  \label{eq:u}
  u\equiv\frac{1}{1-\epsilon_B/\epsilon_A}
\end{equation}
se conoce como la {\em variable espectral}, y
\begin{equation}
  \label{eq:B}
  B(\bm r)=\left\{
      \begin{matrix}
        0,\quad\text{si $\bm r\in A$}\\
        1,\quad\text{si $\bm r\in B$}
      \end{matrix}
    \right.
\end{equation}
es la {\em función característica}. Notemos que $u$ depende de la
composición del material y puede depender de la frecuencia a través de
las funciones dieléctricas de las componentes, mientras que $B(\bm r)$ no
depende de la composición ni de la frecuencia, sino únicamente de la
geometría.

Supongamos que nuestro sistema es periódico, caracterizado por
una {\em red de Bravais} $\{\bm R\}$, tal que $\epsilon(\bm r+\bm
R)=\epsilon(\bm r)$. A esta red le corresponde una {\em red recíproca}
$\{\bm K\}$, formada por todos aquellos vectores de onda tales que el
producto escalar $\bm K\cdot\bm R=2\pi n$, con $n$ un número entero
para cualquier vector real $\bm R$ y vector recíproco $\bm K$.

Entonces, podemos describir a $\hat \epsilon$
como una matriz en el espacio recíproco mediante una integral de Fourier,
\begin{equation}
  \label{eq:eKK}
  \epsilon_{\bm K\bm K'}=\epsilon_{\bm K-\bm K'}=\frac{1}{\Omega}\int_0
  d^3r\, \epsilon(\bm r) e^{-i\bm K\cdot\bm r}.
\end{equation}
donde la integral se realiza sobre una celda primitiva cualquiera, cuyo
volumen es $\Omega$. La relación conversa es
\begin{equation}
  \label{eq:er}
  \epsilon(\bm r)=\sum_{\bm K}\epsilon_{\bm K} e^{i\bm K\cdot\bm r}.
\end{equation}
El teorema de Bloch nos permite escribir a los campos en el interior
de un sistema periódico como una superposición de ondas de Bloch, cada
una de las cuales cambia por una fase $e^{i\bm k\cdot\bm R}$ cuando
nos desplazamos un vector de la red, donde $\bm k$ es el {\em vector
  de Bloch}, una especie de vector de onda, que es una {\em cantidad
  conservada}. Por tanto, eligiendo un valor cualquiera de $\bm k$
podemos escribir cualquier campo en términos de una suma de Fourier
como
\begin{equation}
  \label{eq:blochFourier}
  \bm F(\bm r)=\sum_{\bm K}\bm F_{K} e^{i(\bm k+\bm K)\cdot\bm r}.
\end{equation}
El {\em teorema de convolución} nos permite entonces escribir la
ecuación material $\bm D(\bm r)=\epsilon(\bm r)\bm E(\bm r)$ en el
espacio recíproco como la ecuación matricial
\begin{equation}
  \label{eq:DKvsEK}
  \bm D_{\bm K}=\sum_{\bm K'}\epsilon_{\bm K\bm K'} \bm E_{\bm K'}.
\end{equation}
Como la respuesta del material puede representarse en distintos
espacios, como son el espacio real o el espacio recíproco,
conviene abstraer la acción de la respuesta dieléctrica y tratarla
como un operador $\hat{\bm \epsilon}$ abstracto, más que como una
función de la posición.

En el espacio recíproco, el operador $\nabla$ se puede representar por
un producto con el vector $\nabla\to i(\bm k+\bm K)$, por lo cual el laplaciano
es $\nabla^2\to-|\bm k+\bm K|^2$ y su inverso es simplemente
$\nabla^{-2}\to -1/|\bm k+\bm K|^2$. Por lo tanto, el proyector
longitudinal \eqref{eq:PL} se puede representar por la matriz
\begin{equation}
  \label{eq:PLKK}
  \mathcal P^L_{\bm K\bm K'}=\delta_{\bm K\bm K'}\hat{\bm K}\hat{\bm K}\cdot,
\end{equation}
donde definimos los vectores unitarios
\begin{equation}
  \label{eq:hatK}
  \hat{\bm K}\equiv\frac{\bm k+\bm K}{|\bm k+\bm K|},
\end{equation}
y $\delta_{\bm K\bm K'}$ es una función delta de Kronecker.

Los vectores recíprocos $\bm K\ne0$ corresponden a oscilaciones
con longitudes de onda del orden de el parámetro de red del sistema
periódico. Es conveniente entonces {\em definir} el promedio como
un filtro pasabajos en el espacio recíproco que elimina todos los vectores
recíprocos, exceptuando $\bm K=0$, convirtiendo una onda de Bloch en
una onda plana con vector de onda $\bm k$. Escribimos entonces
\begin{equation}
  \label{eq:PaKK}
  \mathcal P_{p,\bm K\bm K'}=\delta_{\bm K\bm 0}\delta_{\bm K'\bm0}.
\end{equation}

De acuerdo a la ec. \eqref{eq:epsLLM}, podemos hallar la respuesta
macroscópica siguiendo los siguientes pasos:
\begin{enumerate}
\item Expresamos la permitividad como una matriz $\epsilon_{\bm K\bm
    K'}$ en el espacio recíproco haciendo una
  transformación de Fourier (ec. \eqref{eq:eKK}).
\item Tomamos su proyección longitudinal, $\epsilon^{LL}_{\bm K\bm
    k'}$  multiplicando a izquierda y derecha por el proyector
  longitudinal (ec. \eqref{eq:PLKK}).
\item Invertimos la matriz resultante en el subespacio de campos
  vectoriales longitudinales $\left(\epsilon^{LL}_{\bm K\bm K'}\right)^{-1}$.
\item Promediamos la inversa tomando el elemento $\bm K=\bm K'=0$.
\item Interpretamos el resultado como el inverso de la proyección
  longitudinal del tensor dieléctrico macroscópico.
\end{enumerate}
El procedimiento anterior puede resumirse como
$\bm\epsilon^{LL}_M=\hat{\bm k}\epsilon^{LL}_M\hat{\bm k}$, donde la
{\em componente longitudinal} de la respuesta es
\begin{equation}
  \label{eq:kek}
  \epsilon_M^{LL}=\hat{\bm k}\cdot\bm\epsilon_M\cdot\hat{\bm k}=\left(\left.\left(\hat{\bm
    K}\cdot\epsilon_{\bm K\bm K'}\hat{\bm
    K'}\right)^{-1}\right|_{\bm K=\bm K'=\bm 0}\right)^{-1}.
\end{equation}
Repitiendo el cálculo indicado por la ec. \eqref{eq:kek} para
distintas direcciones $\hat{\bm k}$ del vector de onda, podemos hallar
todas las componentes del tensor dieléctrico macroscópico
$\bm\epsilon_M$.
\subsection{Recursión de Haydock}\label{sec:haydock}
De acuerdo a la sección anterior, para obtener la respuesta
macroscópica de un medio binario arbitrario en el límite de longitud
de onda larga, basta invertir la matriz $\hat{\bm K}\cdot\epsilon_{\bm K\bm
  K'}\hat{\bm K'}$ y tomar el elemento $\bm 0\bm 0$ del resultado. Sustituyendo la
ec. \eqref{eq:epsmic1}, obtenemos
\begin{equation}
  \label{eq:evsG}
  \frac{1}{\epsilon_M^{LL}}=\frac{u}{\epsilon_A}\left.(u\delta_{\bm K\bm
    K'}-B^{LL}_{\bm K\bm K'})^{-1}\right|_{\bm K=\bm K'=0},
\end{equation}
donde introdujimos la componente longitudinal de la función
característica en el espacio recíproco
\begin{equation}
  \label{eq:BLL}
  B^{LL}_{\bm K\bm K'}=\hat{\bm K}\cdot B_{\bm K-\bm K'}\hat{\bm K'},
\end{equation}
y $B_{\bm K-\bm K'}$ es el coeficiente de Fourier de la función
característica $B(\bm r)$ correspondiente al vector recíproco $\bm
K-\bm K'$.

Notamos que en la ec. \eqref{eq:evsG} debemos calcular el inverso de un
operador $(u-\hat B^{LL})$ representado como una matriz en el espacio
recíproco, donde $u$ es un número complejo y $\hat B^{LL}$ es un
operador hermitiano, y luego proyectar el resultado sobre un estado
$\ket{0}$ correspondiente a una onda plana con vector de onda $\bm
k$. Esto es análogo al cálculo del operador de Green proyectado
$\braket{0|\hat{\mathcal G}(\varepsilon)|0}$ en mecánica cuántica, donde
$\hat{\mathcal G}(\varepsilon)=(\varepsilon-\hat{\mathcal H})^{-1}$ es el operador
de Green correspondiente a un operador hamiltoniano ${\mathcal H}$
evaluado para un valor $\varepsilon$ de una energía compleja. Entre
las aplicaciones del operador de Green proyectado se encuentra el
cálculo de la densidad de estados cuánticos proyectada. Podemos
entonces tomar prestado el método recursivo de Haydock
\cite{haydock} para calcular proyecciones de funciones de Green. De
acuerdo a nuestra analogía, $u$ juega el papel de
energía compleja $\varepsilon$ y $\hat B^{LL}$ juega el papel de
hamiltoniano $\hat {\mathcal H}$.

Hemos definido el estado $\ket{0}$ como el correspondiente a una onda
plana con vector de onda $\bm k$, i.e., un estado cuya representación
en el espacio recíproco es $\braket{\bm K|0}=\delta_{\bm K\bm 0}$,
pues no tiene contribuciones de ondas con $\bm K\ne0$. Ahora, podemos
generar un nuevo estado haciendo actuar a nuestro hamiltoniano sobre
el estado inicial, $\ket{\tilde 1}=\hat{\mathcal
  H}\ket{0}$. Escribimos este estado como una combinación lineal del
estado que ya teníamos $\ket{0}$ y un estado nuevo $\ket{1}$ del cual
pedimos que sea ortogonal a $\ket{0}$ y que esté normalizado,
$\ket{\tilde 1}=b_1\ket{1}+a_0\ket{0}$. Ahora repetimos el
procedimiento con el estado $\ket{1}$, i.e., $\ket{\tilde
  2}=\hat{\mathcal H}\ket{1}=b_2\ket{2}+a_1\ket{1}+b_1\ket{0}$. El
caso genérico sería
\begin{equation}
  \label{eq:tilden}
  \ket{\tilde n}=\hat{\mathcal
    H}\ket{n-1}=b_n\ket{n}+a_{n-1}\ket{n-1}+b_{n-1}\ket{n-2},
\end{equation}
donde exigimos que todos los estados sean ortonormales, es decir,
\begin{equation}
  \label{eq:n|m}
  \braket{n|m}=\delta_{nm}.
\end{equation}
Aquí empleamos el producto escalar
\begin{equation}
  \label{eq:braket}
  \braket{n|m}=\sum_{\bm K}\phi_n^*(\bm K)\phi_m(\bm K)
  =\frac{1}{\Omega}\int_0 d^3r\, \phi_n^*(\bm r)\phi_m(\bm r)
\end{equation}
donde, copiando el lenguaje de la mecánica cuántica, $\phi_m(\bm
K)=\braket{\bm K|n}$ es la {\em función de onda}
correspondiente al estado $\ket{n}$ evaluada en el vector recíproco
$\bm K$, y $\phi_m(\bm r)=\braket{\bm r|n}$ es la función de onda
correspondiente al mismo estado pero evaluada en la posición $\bm r$,
y donde hemos empleado la identidad de Parseval.

Notamos que en la ec. \eqref{eq:tilden} no aparecen los términos
$\ket{n-3}$, $\ket{n-4}$, etc. pues nuestro operador es
hermitiano. Por ejemplo,
i.e., $\braket{n-3|\tilde
  n}=\braket{n-3|H|n-1} = \braket{n-1|H|n-3}^* =
(\bra{n-1}\left(b_{n-2}\ket{n-2} + a_{n-3}\ket{n-3}+b_{n-3}\ket{n-4}\right))^* = 0$.
Los coeficientes de Haydock son reales y pueden obtenerse de la condición de ortonormalidad,
\begin{equation}
  \label{eq:qn}
  a_{n-1}=\braket{n-1|\hat{\mathcal H}|n-1}
\end{equation}
y
\begin{equation}
  \label{eq:bn}
    \ket{n}=(\ket{\tilde n}-a_{n-1}\ket{n-1}-b_{n-1}\ket{n-1})/b_n,
    \quad \braket{n|n}=1.
\end{equation}
De esta manera podemos construir una {\em base} $\{\ket{n}\}$ en la
cual el {\em hamiltoniano} $\hat{\mathcal H}$ puede representarse por
una matriz tridiagonal
\begin{equation}
  \label{eq:trid}
  \mathcal H_{nn'}=
  \begin{pmatrix}
    a_0   &b_1   &0     &0     &0\ldots\\
    b_1   &a_1   &b_2   &0     &0\ldots\\
    0     &b_2   &a_2   &b_3   &0\ldots\\
    0     &0     &b_3   &a_3   &b_4\ldots\\
    \vdots&\vdots&\vdots&\ddots&\ddots\ddots\\
  \end{pmatrix}
\end{equation}
De acuerdo a la ec. \eqref{eq:evsG}, la respuesta macroscópica está
dada por la proyección de la inversa $\mathcal G_{nn'}$ de la matriz tridiagonal
$\varepsilon-\mathcal H_{nn'}$ sobre el estado $\ket{0}$, i.e., no
necesitamos toda la inversa, sino sólo su elemento $\mathcal G_{00}$.
Notamos que el vector columna $\mathcal G_{n0}$ obedece la ecuación
\begin{equation}
  \label{eq:Gn0}
  \sum_{m} (\varepsilon\delta_{nm}-\mathcal H_{nm})\mathcal G_{m0}=\delta_{n0}.
\end{equation}
Si truncamos la ecuación después de $N$ renglones, el último renglón
de esta ecuación sería de la forma
\begin{equation}
  \label{eq:GN}
  -b_N\mathcal G_{N-1,0} +(u-a_N)\mathcal G_{N,0}=0,
\end{equation}
lo cual nos permite despejar
\begin{equation}
  \label{eq:GN1}
  \mathcal G_{N,0}=\frac{b_N}{u-a_N}\mathcal G_{N-1,0}.
\end{equation}
Sustituyendo esta solución en la penúltima ecuación
\begin{equation}
  \label{eq:GN2}
  \begin{split}
    -b_{N-1}\mathcal \mathcal_{N-2,0}
    &+(u-a_{N-1})\mathcal G_{N-1,0}-b_{N}\mathcal
    G_{N,0}\\
    &=-b_{N-1}\mathcal G_{N-2,0}+
    \left(u-a_{N-1}-\frac{b_N^2}{u-a_N}\right)\mathcal G_{N-1}=0,
  \end{split}
\end{equation}
podemos despejar
\begin{equation}
  \label{eq:GN3}
  \mathcal G_{N-1,0}=\frac{b_{N-1}}{u-a_{N-1}-\frac{b_N^2}{u-a_N}}\mathcal G_{N-2,0}.
\end{equation}
Prosiguiendo de esta forma con todas las ecuaciones correspondientes a $n>0$,
\begin{equation}
  \label{eq:Gnne0}
  -b_n\mathcal G_{n-1,0} +(u-a_n)\mathcal G_{n,0}-b_{n+1}\mathcal G_{n+1,0}=0,
\end{equation}
llegamos a
\begin{equation}
  \label{eq:GN4}
  \mathcal
  G_{1,0}=\frac{b_{1}}{u-a_{1}-\frac{b_2^2}{u-a_2-\frac{b_3^2}{u-a_3-\ddots}}}
  \mathcal G_{0,0}.
\end{equation}
Sustituyendo en la ecuación correspondiente a $n=0$,
\begin{equation}
  \label{eq:Gneq0}
  (u-a_0)\mathcal G_{00}-b_{n+1}\mathcal G_{1,0}=1,
\end{equation}
obtenemos una expresión para $\mathcal G_{00}$ en forma de una
fracción continuada
\begin{equation}
  \label{eq:G00}
  \mathcal
  G_{00}=\frac{1}{u-a_0-\frac{b_{1}^2}{u-a_{1}-\frac{b_2^2}{u-a_2-\frac{b_3^2}{u-a_3-\ddots}}}},
\end{equation}
la cual podemos emplear en la ec. \eqref{eq:evsG} para finalmente
obtener la respuesta macroscópica
\begin{equation}
  \label{eq:eMLL}
  \epsilon_M^{LL}=\frac{\epsilon_A}{u}\left(
    u-a_0-\frac{b_{1}^2}{u-a_{1}-\frac{b_2^2}{u-a_2-\frac{b_3^2}{u-a_3-\ddots}}}\right).
\end{equation}

Debemos enfatizar que en esta expresión, los coeficientes de Haydock
$\{a_n, b_n\}$ dependen exclusivamente de la geometría a través de la función
característica $B(r)$ y no de la composición del material ni
de la frecuencia. Por lo tanto, sólo es necesario calcularlos una vez
para una geometría dada y posteriormente pueden emplearse
para calcular la respuesta de cualquier metamaterial con dicha
geometría con cualquier composición y a cualquier frecuencia,
simplemente sustituyendo la variable espectral adecuada $u$
(ec. \eqref{eq:u}). Así, con este formalismo podemos calcular la
respuesta de sistemas formados por aislantes o metales, con o sin
dispersión y con o sin disipación; el operador $\hat B^{LL}$ es
hermitiano haya o no haya dispersión o disipación.

Por otro lado, podemos  aplicar el hamiltoniano $\hat
B^{LL}$ \eqref{eq:BLL} en
etapas, notando que multiplicar por $\hat{\bm K'}$ es trivial en el
espacio recíproco. El producto matricial del vector resultante con
$B_{\bm K\bm K'}=B_{\bm K-\bm K'}$ corresponde a una convolución, por lo
cual, tras una transformada de Fourier hacia el espacio real, se
convierte  en un producto trivial por la función
característica $B(\bm r)$. Finalmente, tomando una transformada de
Fourier de regreso al espacio recíproco, el producto escalar por los vectores
unitarios $\hat{\bm K}\cdot$ se vuelve trivial. Esto muestra que podemos
aplicar nuestro hamiltoniano repetidas veces para obtener los
coeficientes de Haydock sin necesidad de multiplicar ninguna
matriz. Eso vuelve muy eficiente el proceso aquí descrito.

\section{Implementación}
La teoría mostrada arriba ha sido implementada en un paquete
computacional llamado {\em Photonic}, el cual ha sido colocado en el
{\em dominio público} \cite{PhotonicCPAN,PhotonicGitHub}. El programa está
escrito en el lenguaje {\em PERL}, el cual es muy expresivo y cuya
sintaxis hereda construcciones de lenguajes previos como {\em C, C++,
  awk, bash, sed}, etc. El lenguaje es versátil y flexible, y refleja
la filosofía de su creador resumida en frases como {\em hay más de una
  manera de resolverlo} y {\em lo fácil debe ser fácil, lo difícil
  debe ser posible}. Sólo contiene tres tipos de datos, que son {\em
  escalares, arreglos} indexados por un entero y {\em arreglos
  asociativos} indexados por cualquier escalar. Los escalares pueden
representar números enteros o reales, cadenas de caracteres, o
referencias a otros escalares, arreglos o arreglos asociativos, e
incluso, referencias a subrutinas. Los arreglos son dinámicos y pueden
crecer o decrecer en ambos extremos o en su interior. Esto permite
implementar de manera trivial pilas de datos ({\em stacks}), colas {\em
  fifo}, árboles y otras estructuras de datos. Se pueden
construir fragmentos de código durante la ejecución de un programa
para ejecutarse posteriormente. Esta flexibilidad permite
emplear una gran variedad de paradigmas al programar en {\em PERL},
incluyendo programación procedural, funcional y/o orientada a
objetos.

El costo a pagar por la flexibilidad del lenguaje es la velocidad de
ejecución. Se ha reportado recientemente que {\em PERL} es varias decenas de
veces más lento que {\em C} para tareas orientadas a procesamiento
numérico.\cite{benchmark} Por ello, un grupo de investigadores se
abocaron a crear una extensión del lenguaje llamado {\em Perl Data
  Language} o {\em PDL} \cite{PDL, PDLCPAN, PDLGitHUb}, que permite
ligar rutinas numéricas escritas en otros lenguajes como {\em C} o
{\em Fortran} para el manejo eficiente de arreglos numéricos, sin
sacrificar la flexibilidad y expresividad de {\em PERL}. Una prueba
reciente\cite{benchmark} mostró que {\em PDL} es competitivo y hasta
puede superar en velocidad a códigos nativos en {\em C}.

Finalmente, para simplificar el proceso de codificación y volverlo
robusto conforme evoluciona el paquete se empleó un sistema de
programación de objetos conocido como {\em Moose}\cite{Moose}. Este
sistema permite definir {\em clases} que abstraen el comportamiento de
los {\em objetos}, instancias que tienen una serie de
{\em atributos}, datos privados, y {\em métodos} que definen su
comportamiento. Las clases pueden heredar su
comportamiento de otras clases o de {\em roles} que definen las
interfases. Contar con una librería de clases permite armar programas
que resuelven problemas complejos juntando bloques que ensamblan unos
con otros, como las construcciones con bloques de juguetes  {\em Lego}.

\subsection{Ejemplo}

No explicaremos aquí los detalles de la implementación, pues su
comprensión requeriría cierto dominio de los sistemas (Perl, PDL y
Moose) empleados. En cambio, mostraremos fragmentos de un pequeño
programa para explicar cómo se usa el sistema. El programa íntegro,
disponible en la referencia \cite{toroide}, calcula
el tensor dieléctrico de una red tetragonal de toroides hechos de cierto
material y embebido en una matriz de otro material.

Iniciamos con una serie de {\em pragmas} y cargando paquetes que serán
útiles más adelante
\begin{verbatim}
#!/usr/bin/env perl
# ...
use strict;
use warnings;
use v5.12;
use Getopt::Long;
use PDL;
use PDL::NiceSlice;
use PDL::Constants qw(PI);
use Photonic::Geometry::FromB;
use Photonic::LE::NR2::Haydock;
use Photonic::LE::NR2::EpsL;
\end{verbatim}
{\tt strict} y {\tt warnings}  son para pedir al sistema que sea estricto y nos advierta
de errores potenciales, {\tt v5.12} es para habilitar algunas
construcciones semánticas, {\tt Getopt} es para leer los parámetros
desde la línea de comandos al ejecutar el programa, {\tt PDL} es para usar la interface
numérica, {\tt NiceSlice} para simplificar el manejo de índices en las
estructuras de datos y {\tt PI} es simplemente una constante
útil. Los paquetes relacionados con {\tt Photonic} serán discutidos
más abajo. Las componentes {\tt LE} y {\tt NR2} en su nombre indican que usaremos
aquellas rutinas relacionadas con la respuesta dieléctrica
longitudinal en el límite no retardado y restringido a dos componentes.

A continuación definimos algunos parámetros y el código para leerlos
desde la línea de comandos.
\begin{verbatim}
my $ratio; # b/a for torus
...
my $options=q(
	'ratio=f'=>\$ratio,
        ...
	);
...
GetOptions( %options)or usage($options, "Bad options");
usage($options, "Missing options")
     unless luall {defined $_}
     ($ratio, $fraction, @eps_a, @eps_b, $Nxy, $Nz, $Nh);
...
set_autopthread_targ($cores) if defined $cores;;
\end{verbatim}
Los parámetros a leer son la razón de los radios del toroide, la
fracción de llenado en la celda unitaria, el numero de {\em voxels} a lo
largo de los ejes de la red, los pares de funciones dieléctricas e
emplear para los toroides y la matriz, el número de coeficientes de Haydock a
emplear y el número de núcleos computacionales a emplear en el
cálculo. A continuación se leen y validan las opciones y de ser
necesario se envían mensajes de error. La rutina {\tt
  set\_autopthread\_targ} establece el número de núcleos computacionales
que deseamos usar al paralelizar el programa.

Como indicamos arriba, la geometría queda definida a partir de la
función característica, cuyo valor es 1 dentro del toroide y 0 en
su exterior. Primero calculamos los dos radios del toroide en términos
de la fracción de llenado deseada.
\begin{verbatim}
my ($Nxy2, $Nz2)=(2*$Nxy+1, 2*$Nz+1);
my $unit_cell_volume=$Nxy2*$Nxy2*$Nz2;
my $small_radius=($fraction*$unit_cell_volume/(2*PI**2*$ratio))**(1/3);
my $large_radius=$ratio*$small_radius;
warn "Tori overlap" if $small_radius>$Nz
                       or $large_radius+$small_radius>$Nxy;
\end{verbatim}
Luego creamos un arreglo 3D representando a la celda unitaria y lo
poblamos de unos y ceros de acuerdo a la función característica
deseada.
\begin{verbatim}
my $r=zeroes($Nxy2, $Nxy2, $Nz2)->ndcoords
      -pdl($Nxy, $Nxy, $Nz); #positions array
my $B=(sqrt($r((0))**2+$r((1))**2)-$large_radius)**2
       +$r((2))**2 < $small_radius**2;
\end{verbatim}
La rutina {\tt zeroes} produce el arreglo 3D de ceros con el número de
dimensiones y el tamaño
solicitado, el {\em método} {\tt ndcoords} asigna a cada punto del
arreglo un vector en 3D con las coordenadas de dicho punto. Al restar
las coordenadas del centro del arreglo, asignamos a la variable {\tt \$r}
las coordenadas de los puntos del arreglo con respecto al
centro. Procesando dichas coordenadas obtenemos la distancia de cada
punto al círculo alrededor del cual se forma el toroide y en la
variable {\tt \$B} guardamos un 1 o un 0 dependiendo de si estamos
suficientemente cerca de la generatriz o no.

A continuación inicializamos dos objetos que codifican la geometría del
sistema empleando la función característica.
\begin{verbatim}
my $gx=Photonic::Geometry::FromB->new(B=>$B, Direction0=>pdl(1,0,0));
my $gz=Photonic::Geometry::FromB->new(B=>$B, Direction0=>pdl(0,0,1));
\end{verbatim}
Las clases {\tt Geometry} saben cómo calcular la red de {\em
  voxels}, pero también cómo calcular la red recíproca, y los
vectores recíprocos normalizados $\hat{\bm K}$, entre
otros métodos relacionados a la geometría del sistema, a partir del
atributo {\tt B} que inicializamos con la
función característica {\tt \$B}. Sin embargo, para ello necesita saber en qué
dirección apunta el vector de Bloch, asociado al atributo {\tt
  Direction0}. Emplearemos entonces dos {\em geometrías},
una con ondas viajando en la dirección $\hat{\bm x}$ y otra viajando en
la dirección $\hat{\bm z}$, i.e., a lo largo del plano y del eje del
toroide, respectivamente.

Con las dos geometrías podemos inicializar dos objetos para calcular
coeficientes de Haydock.
\begin{verbatim}
my $nrx=Photonic::LE::NR2::Haydock->new(geometry=>$gx, nh=>$Nh);
my $nrz=Photonic::LE::NR2::Haydock->new(geometry=>$gz, nh=>$Nh);
\end{verbatim}
Finalmente, para cada pareja de funciones dieléctricas, calculamos la
respuesta dieléctrica macroscópica longitudinal, proyectada sobre la
dirección del vector de Bloch establecida arriba.
\begin{verbatim}
say "#ratio Nxy Nz Nh f-nom f-act medium torus epsxx epszz";
foreach(0..@eps_a-1){
    my ($ea, $eb)=(pdl($eps_a[$_])->r2C, pdl($eps_b[$_])->r2C);
    my $epsx_calc=Photonic::LE::NR2::EpsL->new(haydock=>$nrx, nh=>$Nh,
             epsA=>$ea, epsB=>$eb);
    my $epsz_calc=Photonic::LE::NR2::EpsL->new(haydock=>$nrz, nh=>$Nh,
             epsA=>$ea, epsB=>$eb);
    my $resultx=$epsx_calc->epsL;
    my $resultz=$epsz_calc->epsL;
\end{verbatim}
El objeto {\tt EpsL} sabe calcular la respuesta dieléctrica
longitudinal y se inicializa con un objeto que calcula
coeficientes de Haydock, con el número de coeficientes que se desea
usar y con las funciones dieléctricas de ambas
componentes. Finalmente, el {\em método} {\tt epsL} regresa el {\em
  valor} de la función dieléctrica macroscópica deseada.

El resto del programa simplemente imprime el resultado y mensajes de
error de ser necesario.
\begin{verbatim}
    say sprintf "%.4f %d %d %d %.4f %.4f %.4f %.4f %.4f %.4f",
    $ratio, $Nxy, $Nz, $Nh, $fraction, $gx->f, $ea->re, $eb->re,
    $resultx->re, $resultz->re;
    say "x-no-covergió" unless $epsx_calc->converged;
    say "z-no-covergió" unless $epsz_calc->converged;
}

sub usage {
    ...
}
\end{verbatim}

Podemos correr el programa como en el siguiente ejemplo, en el que
calculamos las propiedades de una red tetragonal de
$161\times161\times41$ voxeles con toroides cuyas funciones
dieléctricas son $\epsilon_b=5,10$, inmersos en el vacío, $\epsilon_a=1$,
con una fracción de llenado nominal $f=0.3$, con una razón entre los
radios del círculo mayor al circulo menor $R_>/R_<=3$ y empleando 100
coeficientes de Haydock, usando en el cálculo 4 núcleos de la unidad
de procesamiento,
\begin{verbatim}
./toroid.pl -ratio 3 -fraction .3 -Nz 20 -Nxy 80 \
    -eps_a 1 -eps_b 5 -eps_a 1 -eps_b 10 -Nh 100 -cores 4
\end{verbatim}
obteniendo unos segundos después la siguiente tabla:
\begin{verbatim}
 #ratio Nxy Nz Nh  f-nom  f-act  medium torus   epsxx  epszz
3.0000  80  20 100 0.3000 0.3004 1.0000 5.0000  1.7228 1.6859
3.0000  80  20 100 0.3000 0.3004 1.0000 10.0000 2.1836 2.0152
\end{verbatim}
Las últimas dos columnas nos proporcionan las componentes del tensor
dieléctrico macroscópico de este sistema.

\subsection{Extensiones}
Además de la teoría no retardada desarrollada en detalle en la sección
\ref{binario}, hemos desarrollado la teoría para poder calcular la
respuesta dieléctrica en presencia de retardamiento (sección
\ref{sec:ret}) y para sistemas no
binarios, con tres o más componentes. También hemos extendido la
teoría para poder calcular los campos electromagnéticos microscópicos
y a partir de ellos calcular propiedades no lineales. Estas
extensiones han sido incorporadas en el paquete {\em Photonic}.

\subsection{Instalación}

Para instalar el paquete {\em Photonic} es necesario instalar primero
el paquete {\em PDL}. En un sistema {\em linux} basta emplear el
comando {\tt cpanm --look PDL} desde una línea de comandos. Debe
leerse entonces el archivo {\tt INSTALLATION} e instalar su lista de
prerrequisitos manualmente para posteriormente invocar los comandos
{\tt perl Makefile.PL}, {\tt make}, {\tt make test} y {\tt make
  install}. Posteriormente, el comando {\tt cpanm Photonic} instala de
manera automática nuestro sistema. Una vez instalado, el comando {\tt
  perldoc Photonic} da acceso al manual en línea.

\section{Resultados}
En esta sección enumeraremos algunos de los resultados que hemos
obtenido con la teoría y códigos descritos arriba.
\subsection{Dicroísmo lineal y transmisión extraordinaria}
\begin{figure}
  \centering
  \includegraphics[width=.8\textwidth]{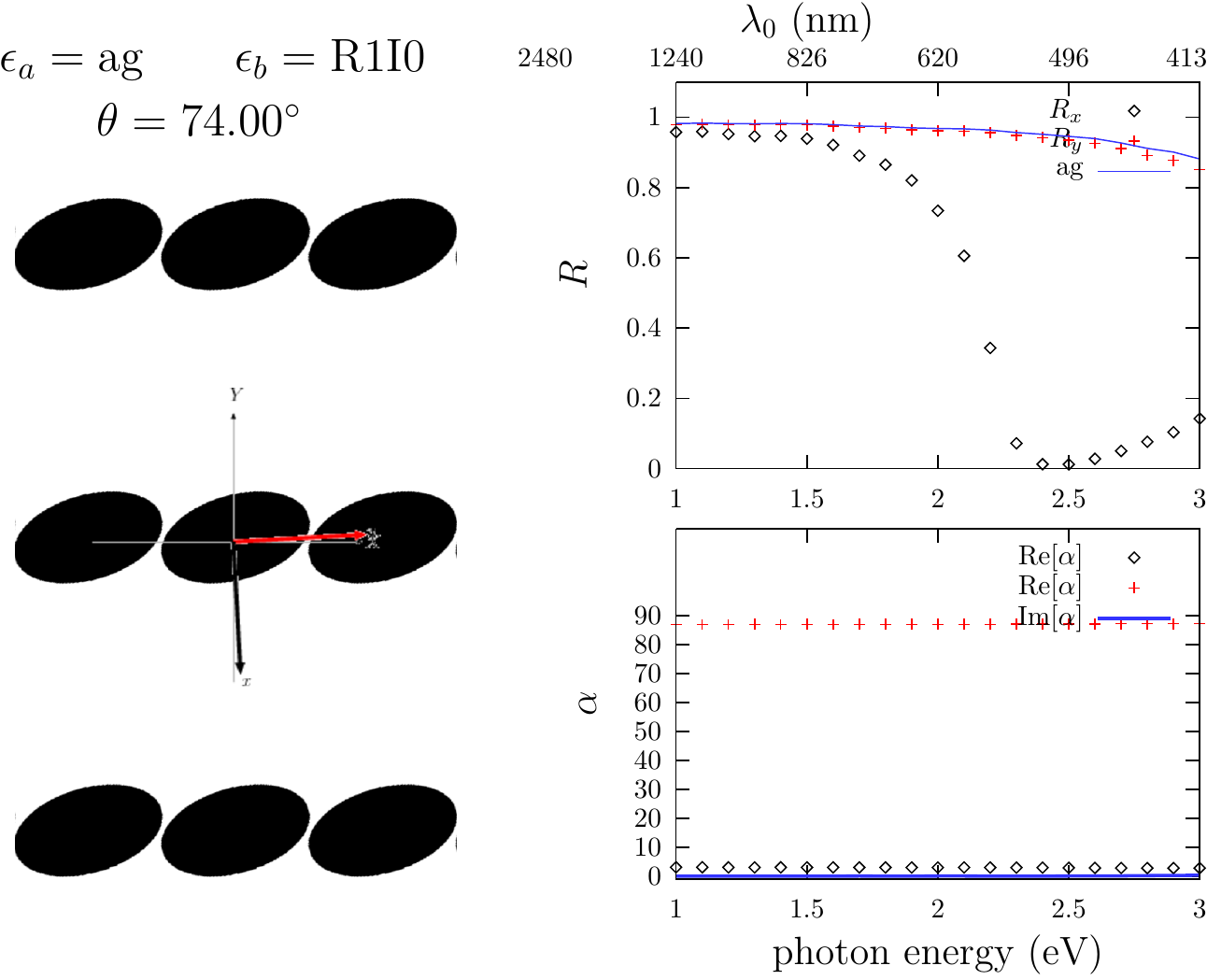}
  \caption{(Izquierda)Red rectangular con parámetros de red $L_y=2L_x$ de
    agujeros cilíndricos con razón de semiejes $a=1.8b$ en una matriz
    de plata. El semieje mayor está rotado $\theta=74^\circ$ con
    respecto a la vertical. (Derecha arriba) Espectro de reflectancia
    de una película delgada de ancho 100\AA iluminada normalmente por
    luz linealmente polarizada a lo largo de los ejes principales del
    tensor dieléctrico, cuya dirección se indica a la izquierda y a la
    derecha abajo. Como referencia, se muestra la reflectancia de una
    película homogénea con la misma cantidad de Ag (linea
    continua). (Derecha abajo) Direcciones principales del tensor
    dieléctrico.}
  \label{fig:elipses}
\end{figure}

El punto de inicio de nuestros cálculos es la función característica
$B(\bm r)$, consistente en unos y ceros, dependiendo de si $\bm r$ se
halla dentro del material $B$ o $A$. Al discretizar el espacio real en
2D,
$B(\bm r)$ se vuelve una representación binaria de una imagen
pixelada. Por lo tanto, nuestro programa puede alimentarse
{\em literalmente} de una imagen pixelada de alto contraste. Esto permite
manipular la imagen usando herramientas gráficas y calcular las
propiedades ópticas del sistema resultante. Ilustramos esto con la
fig. \ref{fig:elipses}, en cuyo lado izquierdo mostramos un corte de
una red rectangular de agujeros cilíndricos con sección transversal
elíptica en una matriz de plata. Manipulamos gráficamente la razón de
aspecto de la red, y la excentricidad y orientación de las elipses. En
la figura hemos elegido el valor 2 para la razón de aspecto de
la red y escogimos 1.8 para la razón entre los semiejes de las
elipses. En el
lado derecho mostramos el espectro de reflectancia a incidencia normal
de una película delgada, de 100\AA\ de ancho, formada por este
metamaterial. Mostramos dos conjuntos de datos, pues el material es
anisótropo. Para cierta polarización, casi horizontal, el material
se comporta como un metal ordinario con una alta reflectancia,
cercana a $R=1$. Sin embargo, para una polarización ortogonal, casi
vertical hay un espectro de reflectancia con un mínimo profundo
alrededor de $\hbar\omega=2.4 eV$, en que nuestra película muestra un
dicroísmo extremo, i.e., casi toda la luz se refleja para una
polarización y casi nada para la polarización ortogonal.\cite{bms12}
El mínimo de reflectancia puede entonarse a través de toda la región
visible mediante cambios pequeños en la orientación $\theta$ de las
elipses. El motivo del dicroísmo extremo es que en la dirección
vertical los caminos conductores se hallan casi estrangulados, pero
abiertos. Por tanto, mientras que el sistema es un buen conductor para
campos horizontales, es un mal conductor para campos verticales y para
campos que oscilan rápidamente el sistema se comporta como un
dieléctrico con resonancias relacionadas a los plasmones localizados
en los cilindros. Por tanto, la permitividad en la dirección vertical
es negativa a bajas frecuencias (comportamiento metálico) pero
positiva y con resonancias a altas frecuencias (comportamiento
dieléctrico). Interpolando entre ambos comportamientos, para alguna
frecuencia intermedia la respuesta empata con la del vacío y el
material adquiere una transparencia extraordinaria, lo cual contrasta
con la alta reflectancia para una polarización ortogonal.
\subsection{Dicroísmo circular}
\begin{figure}
  \centering
  \includegraphics[width=.3\textwidth]{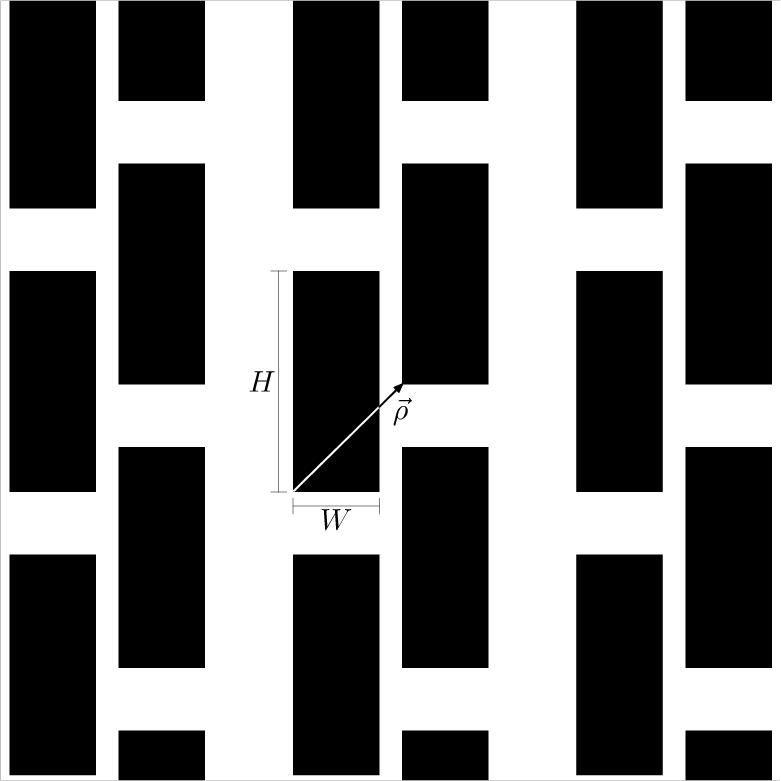}
  \includegraphics[width=.65\textwidth]{fig16b}
  \caption{(Izquierda) Película de Ag  depositada en vidrio con una
    red cuadrada de parejas de agujeros en forma de prismas
    rectangulares rellenos de un dieléctrico con permitividad
    $\epsilon_b$. El sistema está
    parametrizado por el ancho $W$ y altura $H$ de los primas  y el
    desplazamiento ${\bm \rho}$ entre parejas de agujeros en la celda
    unitaria. (Derecha) Dicroísmo circular de películas de 100nm de
    ancho con parámetros entonados para maximizar el dicroísmo a una
    frecuencia preestablecida (puntos).}
  \label{fig:circular}
\end{figure}

En la fig. \ref{fig:circular} mostramos un sistema formado por una red
cuadrada de parejas de agujeros en forma de prismas rectangulares en una película delgada
de Ag colocada sobre un sustrato de vidrio. Los agujeros están
rellenos de un dieléctrico con permitividad $\epsilon_b$. El sistema
está parametrizado por el ancho $W$ y alto $H$ de los prismas y por el
desplazamiento relativo $\bm \rho$ entre cada pareja. Notamos que
este sistema no tiene simetría de reflexión a lo largo del plano,
excepto para ciertos valores particulares del desplazamiento y que,
aunque simétrica, su respuesta dieléctrica macroscópica es no
hermitiana. Por lo tanto, los ejes principales en los que se
diagonaliza el tensor dieléctrico de esta estructura son en general
complejos, los modos propios respectivos corresponden a polarización elíptica y
dependen en general de la frecuencia. Podemos aprovechar estas
características del sistema para diseñar varios
dispositivos ópticos. Como el cálculo reseñado en la
sec. \ref{sec:haydock} es muy eficiente, podemos calcular espectros completos para
cada una de las combinaciones de parámetros que surjan en una búsqueda
automatizada del óptimo de cualquier propiedad deseada.
\cite{bms16}
Por ejemplo, hemos hallado las combinaciones de parámetros que nos permiten obtener
un máximo de dicroísmo circular, la diferencia entre la absortancia de
la película cuando es iluminada con polarización circular derecha e
izquierda, y situarlo en cualquier frecuencia
deseada.\cite{bms16} En el lado derecho de la fig. \ref{fig:circular}
mostramos fragmentos de los espectros de dicroísmo obtenidos,
mostrando que podemos entonar su máximo a cualquier energía deseada
en el espectro visible. Hacemos notar que el dicroísmo circular de los
materiales naturales suele ser de apenas unas partes en mil, mientras
que aquí hemos encontrado señales de orden uno.
\subsection{Magnetismo}
La permitividad obtenida de acuerdo al procedimiento descrito en la
sección \ref{sec:noret} conduce a una permitividad que en general es
{\em no-local}, es decir, $\bm\epsilon(\bm k,\omega)$ depende
explícitamente del
vector de onda $\bm k$ además de depender de la frecuencia
$\omega$. Una de las consecuencias esta no-localidad o {\em dispersión
  espacial}, es que la
permitividad incluye información sobre la respuesta magnética del
sistema. Para entender cómo un sistema no magnético adquiere
propiedades magnéticas cuando se excita con un campo eléctrico cuya
longitud de onda es del orden de las otras escalas de distancia del
sistema, consideremos un cilindro metálico. Si iluminamos el cilindro
con un campo cuya longitud de onda fuese el doble del diámetro del
cilindro, entonces el campo eléctrico induciría corrientes en una
dirección en la mitad del cilindro y en la dirección opuesta en la
otra mitad, i.e., induciría una corriente que circularía alrededor del
cilindro, generando un dipolo magnético de origen eléctrico.

\begin{figure}
  \centering
  \includegraphics[width=.34\textwidth]{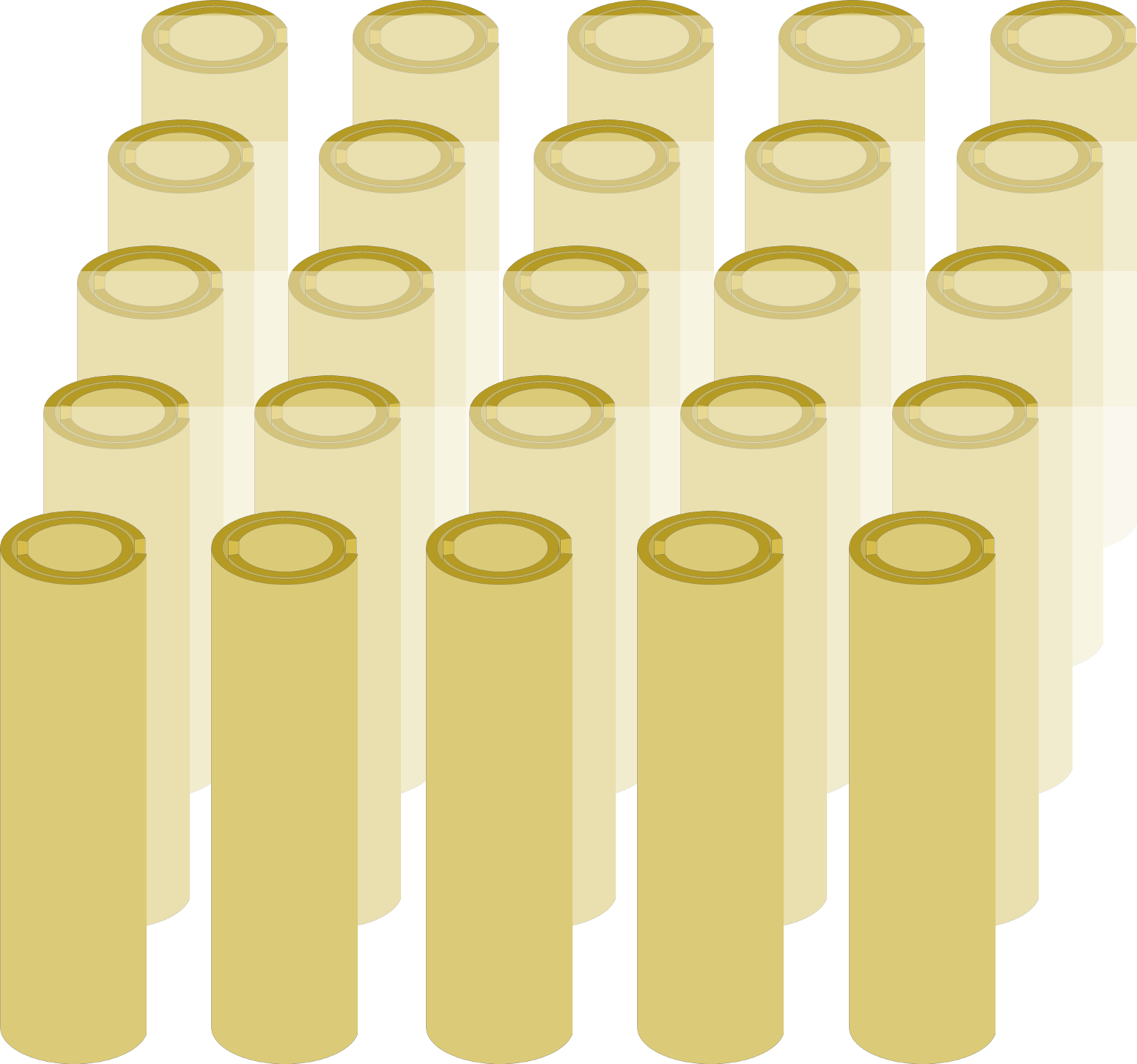}
  \includegraphics[width=.55\textwidth]{fig17b}
  \caption{(Izquierda) Red cuadrada de parejas de cilindros metálicos cortados
    (split rings) anidados. Los cilindros están descritos por el
    modelo de Drude con frecuencia de plasma $\omega_p=20c/d$, con $d$
    el parámetro de red. (Derecha) Permeabilidad magnética obtenida
    a partir de la dispersión espacial de la permitividad y a partir de
    un modelo simple para cilindros con paredes infinitesimales pero con
    la misma conductividad superficial. Los radios internos y externos
    de los anillos metálicos son $0.26d$, $0.32d$, $0.34d$, y
    $0.4d$ y están interrumpidos por brechas de tamaño $0.1d$. Se
    emplearon 350 pares de coeficientes de Haydock en una retícula de
    $401\times401$ pixeles. }
  \label{fig:mu}
\end{figure}

Un procedimiento simple para extraer la respuesta magnética a partir
de la dispersión espacial de la respuesta dieléctrica consiste en
analizar la relación de dispersión de las ondas electromagnéticas en
el medio no local,
\begin{equation}
  \label{eq:kvsw}
  k^2=\epsilon_M(k,\omega) \frac{\omega^2}{c^2},
\end{equation}
donde por simplicidad ignoramos el carácter tensorial de $\epsilon$ y
vectorial de $k$. Hacemos una expansión de Taylor
\begin{equation}
  \label{eq:taylor}
  \epsilon_M(k,\omega)=\epsilon_M(0,\omega)+\frac{k^2}{2}\frac{\partial^2}{\partial
    k^2}\epsilon_M(0,\omega)+\ldots,
\end{equation}
respecto al número de onda para $k$'s pequeñas. El término lineal
está ausente de esta expansión si el sistema es invariante frente a
inversiones temporales. Sustitución en la ec. \eqref{eq:kvsw} conduce
aproximadamente a
\begin{equation}
  \label{eq:kvsw1}
  k^2= \left(\epsilon_M(0,\omega)+\frac{k^2}{2}\frac{\partial^2}{\partial
    k^2}\epsilon_M(0,\omega)\right) \frac{\omega^2}{c^2}.
\end{equation}
Despejando $k^2$ obtenemos
\begin{equation}
  \label{eq:kvsw2}
  k^2= \frac{\epsilon_M(0,\omega)}{1-\frac{\omega^2}{2c^2}\frac{\partial^2}{\partial
    k^2}\epsilon_M(0,\omega)} \frac{\omega^2}{c^2},
\end{equation}
la cual podemos escribir como
\begin{equation}
  \label{eq:kvsw3}
  k^2=\epsilon_M(\omega)\mu_M(\omega)\frac{\omega^2}{c^2},
\end{equation}
donde definimos $\epsilon_M(\omega)$ como el límite local
$\epsilon_M(k\to0, \omega)$ de la permitividad no local, y donde
identificamos la permeabilidad local
\begin{equation}
  \label{eq:mu}
  \mu_M(\omega)=\frac{1}{1-\frac{\omega^2}{2c^2}\frac{\partial^2}{\partial
    k^2}\epsilon_M(0,\omega)}.
\end{equation}
En la fig. \ref{fig:mu} mostramos un sistema formado por una red
cuadrada, de anillos concéntricos truncados. Calculamos con {\em
  Photonic} la permeabilidad no local del sistema, y a partir de su
dispersión espacial obtuvimos la permeabilidad magnética.\cite{lucila} Del lado
derecho mostramos el espectro de la permeabilidad $\mu_M(\omega)$ como
función de lq frecuencia normalizada $qd=\omega d/c$ para un sistema de
anillos metálicos, descritos por una respuesta de Drude con frecuencia
de plasma $\omega_p=20c/d$, con $d$ el parámetro de red. Los radios
internos y externos de los anillos son $0.26d$, $0.32d$, $0.34d$, y
$0.4d$ y están interrumpidos por brechas de tamaño $0.1d$. Los
cálculos fueron realizados en una retícula de $401\times401$ pixeles y
se emplearon 350 pares de coeficientes de Haydock.
Como referencia, se muestran resultados de un cálculo simplificado
para una red de anillos infinitamente delgados pero con una
conductividad superficial que corresponde a los anillos sólidos. Hay
un buen acuerdo entre ambos resultados. Se
observa que alrededor de $qd=0.3$ aparece una resonancia en la
permeabilidad, arriba de la cual adquiere valores negativos, por lo
cual este sistema podría emplearse para construir un metamaterial
izquierdo.

\subsection{Respuesta no lineal}

\begin{figure}
  \centering
  \includegraphics[width=.7\textwidth]{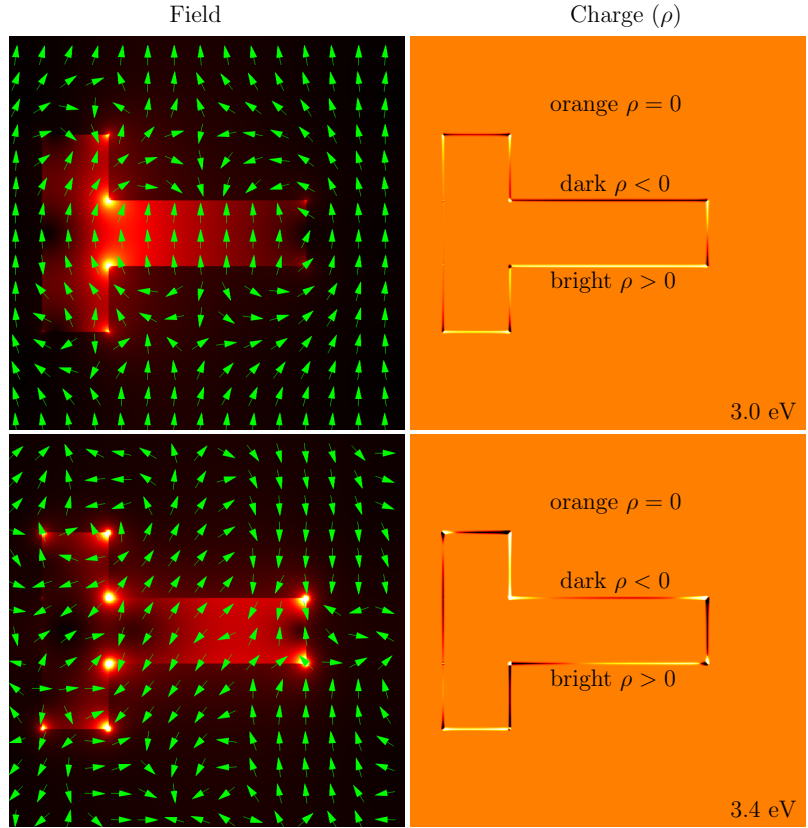}
  \caption{Magnitud y dirección del campo lineal microscópico
    (izquierda) y la densidad de carga inducida (derecha) en un
    metamaterial formado por una red de agujeros en forma de letra T
    en una matriz de Ag excitado por un campo polarizado en la
    dirección vertical para dos frecuencias resonantes. El carácter
    dipolar y cuadrupolar de las resonancias se advierte en los signos
    de la carga inducida y las direcciones del campo.}
  \label{fig:lineal}
\end{figure}

\begin{figure}
  \centering
  \includegraphics[width=\textwidth]{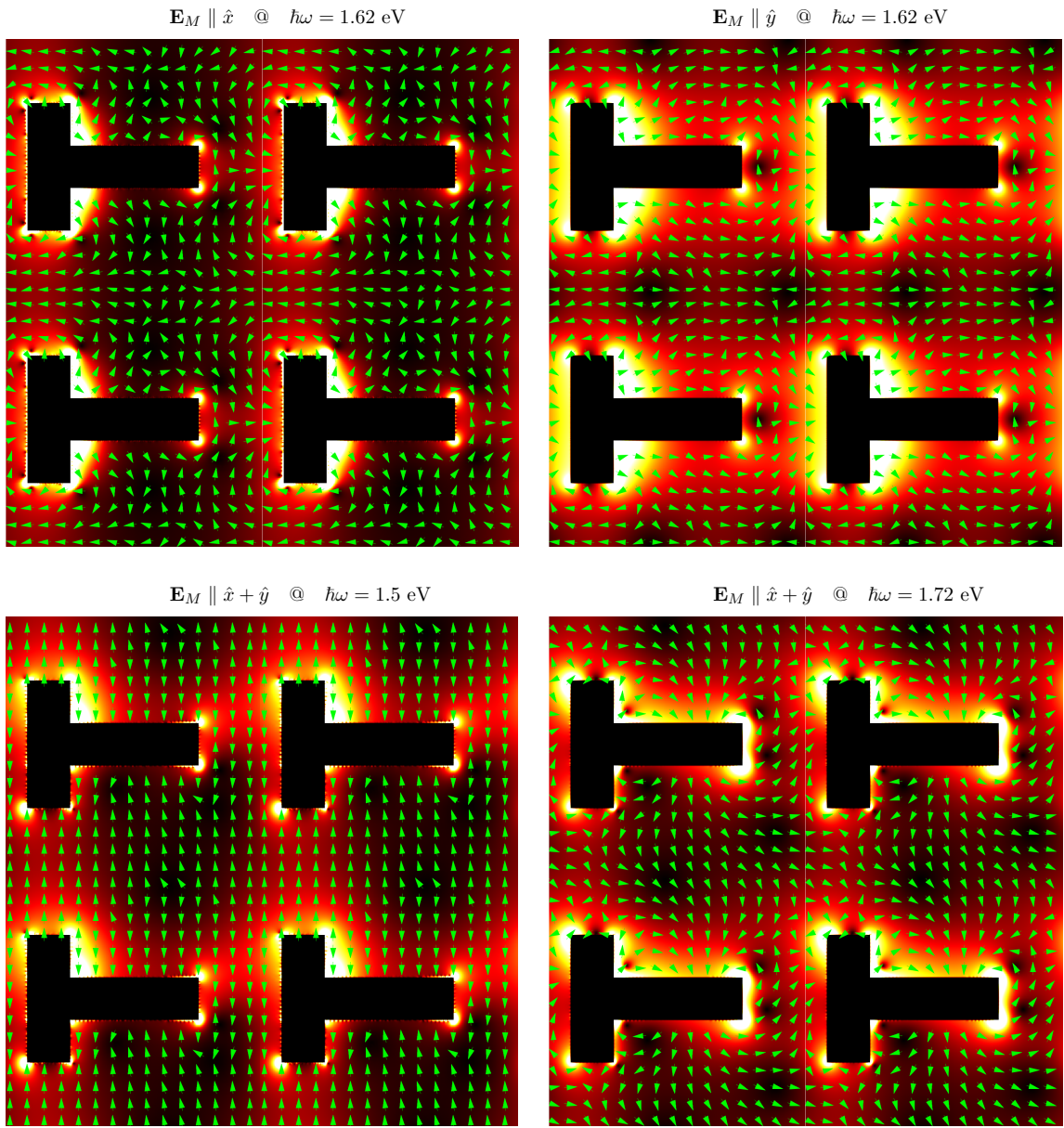}
  \caption{Magnitud y dirección de la polarización cuadrática no
    lineal inducida en el segundo armónico para el mismo sistema
    descrito en la fig. \ref{fig:lineal} para diversas frecuencias y
    diversas direcciones del campo eléctrico macroscópico lineal.}
  \label{fig:P2}
\end{figure}

Además de obtener la respuesta macroscópica, es posible obtener con
una ligera extensión del formalismo presentado arriba el campo
eléctrico microscópico en el seno de un metamaterial. El campo
microscópico permite hacer cálculos de propiedades no lineales, tales
y como la generación de segundo armónico proporcional al cuadrado del
campo. En sistemas
centrosimétricos, aquellos con simetría de inversión, no se pueden
llevar a cabo procesos cuadráticos, en los que se absorben dos fotones
y se emite un fotón con la suma de sus energías. En particular, no se
pueden llevar a cabo procesos de generación de segundo armónico, en
que dos fotones de frecuencia $\omega$ se combinen entre sí para dar
lugar a un fotón de frecuencia $2\omega$. Sin embargo, en la vecindad
de superficies estos procesos sí están permitidos, aunque para
superficies centrosimétricas, las contribuciones de distintas partes
opuestas de la superficie se cancelan mutuamente. Por esto es
interesante calcular la respuesta no lineal de metamateriales formados
por materiales centrosimétricos pero con geometrías no
centrosimétricas. Como un ejemplo,\cite{bms19} en la fig. \ref{fig:lineal}
mostramos el campo lineal microscópico y la densidad de carga inducida
en una red de agujeros no centrosimétricos con forma de letra T en el
seno de una película de Ag.
Debido a la no homogeneidad del campo lineal, en este sistema se
induce una polarización cuadrática que oscila en el segundo armónico,
como ilustra la fig. \ref{fig:P2} para diversas direcciones de
polarización del campo lineal macroscópico y diferentes frecuencias.
\begin{figure}
  \centering
  \includegraphics[width=\textwidth]{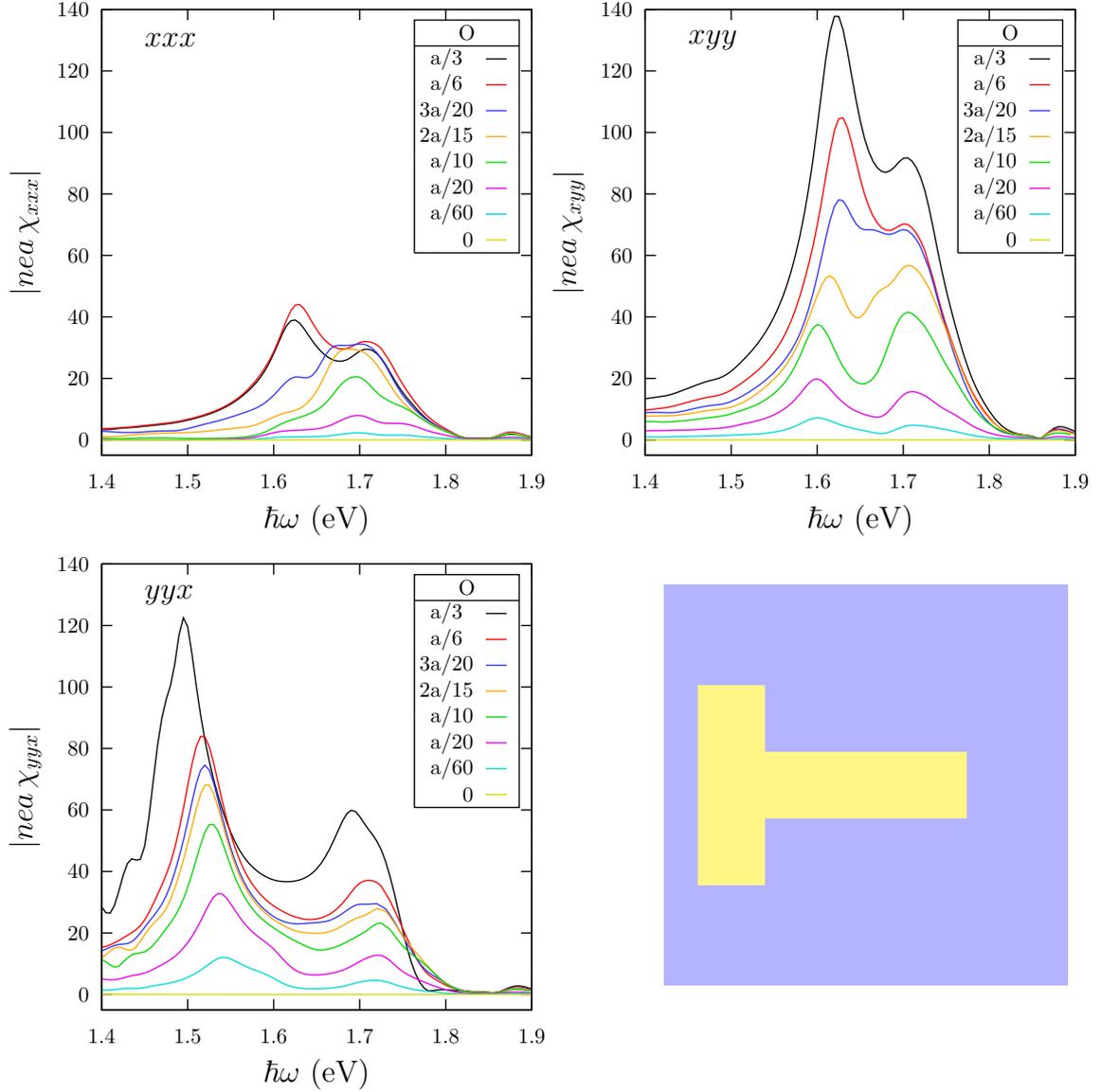}
  \caption{Distintas componentes de la susceptibilidad no lineal para
    generación de segundo armónico $\chi_{ijk}$ para el sistema
    descrito en la figura \ref{fig:lineal} como función de la energía
    de los fotones fundamentales. Distintas curvas corresponden a las
    contribuciones de zonas a distintas distancias de la superficie.}
  \label{fig:xi2}
\end{figure}
Notamos que los patrones no lineales son simétricos, como el sistema,
ante una reflexión $y\leftrightarrow -y$ cuando el campo $\bm E_M$
apunta en las direcciones $\hat{\bm x}$ o $\hat{\bm y}$, pero que esta simetría se pierde
cuando el campo apunta en otras direcciones como la $\hat{\bm
  x}+\hat{\bm y}$.
A partir de la polarización no lineal podemos calcular todas las
componentes del tensor de susceptibilidad no lineal del metamaterial y
podemos optimizarlo a través de los parámetros geométricos del
sistema. En la fig. \ref{fig:xi2} mostramos las componentes
$\chi_{ijk}(\omega,\omega;2\omega)$ no nulas de la susceptibilidad
cuadrática para la generación de segundo armónico en el sistema. La
susceptibilidad típica de un material no centrosimétrico es del orden
de $1/nea_B$, donde $n$ es la densidad de número atómica, $e$ la carga
del electrón y $a_B$ el radio de Bohr. Nuestros resultados muestran
que un metamaterial hecho de componentes centrosimétricas pero con una
geometría no centrosimétrica puede alcanzar en resonancia
susceptibilidades del orden de $100/nea$, con $a$ el parámetro de
red. Por lo tanto, para materiales nanoestructurados, la respuesta no
lineal de nuestros metamateriales puede ser competitiva con la de los
materiales no lineales usuales.

\section{Conclusiones}
En este trabajo hemos presentado una introducción a los metamateriales
y a algunas de sus múltiples propiedades, algunas exóticas, y
aplicaciones. Luego desarrollamos una teoría basada en la
identificación de operadores cuyo promedio tiene significado físico y
a partir de los cuales podemos obtener las funciones respuesta
macroscópicas del sistema y sus propiedades ópticas. También
presentamos el método recursivo de Haydock, el cual aprovecha una
analogía entre el cálculo de funciones respuesta macroscópicas y
funciones de Green proyectadas para obtener algoritmos computacionales
muy eficientes. Los diversos métodos desarrollados han sido
implementados en paquetes computacionales modulares que han sido
puestos en el dominio público. Como algunos ejemplos de su uso,
presentamos el cálculo, diseño y optimización de propiedades ópticas
lineales y no lineales como son la transmitancia
extraordinaria, el dicroísmo lineal, el dicroísmo circular, las
resonancias magnéticas y la generación de segundo armónico.

\section{Agradecimientos}
Este trabajo fue apoyado por DGAPA-UNAM mediante el proyecto IN111119.
Parte del trabajo aquí reportado fue realizado en colaboración con varios
colegas, incluyendo a Guillermo P. Ortiz, Bernardo S. Mendoza, José
Samuel Pérez-Huerta, Lucila Juárez Reyes y Raksha Singla, así como sus
grupos de trabajo y estudiantes.

\end{document}